\documentclass{article}
\usepackage{natbib,psfig,emulateapj}

\newcommand\nh{\hbox{{$N_{\rm H}$}}}
\newcommand\nhgal{\hbox{{$N_{\rm H}^{\rm Gal}$}}}

\newcommand\mabs{\hbox{{$M_{\rm abs}$}}}
\newcommand\mhot{\hbox{{$M_{\rm hot}$}}}
\newcommand\tacc{\hbox{{$t_{\rm acc}$}}}
\newcommand\msun{\hbox{{$M_{\sun}$}}}
\newcommand\msunyr{\hbox{{$M_{\sun}$ yr$^{-1}$}}}
\newcommand\mdot{\hbox{{$\dot{M}$}}}

\newcommand\fe{\hbox{{$Z_{\rm Fe}$\thinspace}}}
\newcommand\solar{\hbox{{$Z_{\sun}$\thinspace}}}

\newcommand\einstein{{\sl Einstein} }
\newcommand\rosat{{\sl ROSAT} }
\newcommand\asca{{\sl ASCA} }
\newcommand\sax{{\sl SAX} }
\newcommand\chandra{{\sl Chandra} }
\newcommand\xmm{{\sl XMM} }

\newcommand\euve{{\sl EUVE} }

\newcommand\xspec{{\sc xspec} }
\newcommand\mekal{{\sc mekal} }

\newcommand\ctspers{{ct s$^{-1}$\thinspace}}
\newcommand\ergcms{{erg cm$^{-2}$ s$^{-1}$\thinspace}}

\newcommand\emin{\hbox{{$E_{\rm min}$}}}

\newcommand\twarm{\hbox{{$T_{\rm w}$}}}
\newcommand\thot{\hbox{{$T_{\rm h}$}}}

\begin{document} 

\title{Oxygen Absorption in M87: Evidence for a Warm+Hot ISM}

\author{David A. Buote\altaffilmark{1}}

\affil{UCO/Lick Observatory, University of California at Santa Cruz,
Santa Cruz, CA 95064; buote@ucolick.org}

\altaffiltext{1}{Chandra Fellow}

\slugcomment{Accepted for Publication in The Astrophysical Journal}

\begin{abstract}

We present a re-analysis of the \rosat PSPC data within the central
100 kpc of M87 to search for intrinsic oxygen absorption similar to
that recently measured in several galaxies and groups. Since M87 is
the brightest nearby galaxy or cluster possessing an average
temperature ($\sim 2$ keV) within the PSPC bandpass, it is the ideal
target for this study.  Using a spatial-spectral deprojection analysis
we find the strongest evidence to date for intrinsic oxygen absorption
in the hot gas of a galaxy, group, or cluster.  Single-phase plasma
models modified by intervening Galactic absorption cannot fit the
0.2-2.2 keV PSPC data as they under-predict the 0.2-0.4 keV region and
over-predict the 0.5-0.8 keV region where the emission and absorption
residuals are obvious upon visual inspection of the spectral
fits. These absorption and emission features are significant out to
the largest radii investigated.  Since the excess emission between
0.2-0.4 keV rules out intrinsic absorption from cold gas or dust, the
most reasonable model for the excess emission and absorption features
is warm, collisionally ionized gas with a temperature of $\sim 10^6$
K. Simple multiphase models (cooling flow, two phases) modified by
both intervening Galactic absorption and by a single oxygen edge
provide good fits and yield temperatures and Fe abundances of the hot
gas that agree with previous determinations by \asca and \sax.

The multiphase models of M87 inferred from the PSPC can account for
the excess EUV emission observed with \euve and the excess X-ray
absorption inferred from \einstein and \asca data above 0.5 keV.  This
evidence for a multiphase warm+hot ISM in M87 essentially confirms the
original detection by Canizares et al. within the central $\sim
2\arcmin$ using the \einstein FPCS.  Although the total mass of the
warm gas implied by the oxygen absorption is consistent with the
matter deposited by a cooling flow, the suppression of the mass
deposition rate and the distortion of the X-ray isophotes in the
region where the radio emission is most pronounced suggest some
feedback effect from the AGN on the cooling gas.
\end{abstract}


\section{Introduction}
\label{intro}

A contentious issue in X-ray astronomy, and more generally in the
study of galaxy formation and evolution, is whether massive galaxies,
groups, and clusters possess inhomogeneous cooling flows wherein large
quantities of gas have cooled and dropped out of the hot phase
resulting in a spatially extended multiphase medium (e.g., Fabian,
Nulsen, \& Canizares 1984; Fabian 1994; Binney 1996).  In the era
before \chandra and \xmm the instrument with the best combined spatial
resolution, bandwidth, and energy resolution for studies of cooling
flows was the \rosat PSPC. The bandwidth (0.1-2.4 keV) of the PSPC is
especially critical because its sensitivity to energies both above and
below the \ion{O}{1} edge (0.532 keV) has allowed us recently to
detect intrinsic oxygen absorption in cooling flow galaxies and groups
which provides the first direct evidence of a spatially extended
multiphase medium in these systems (Buote 2000b, hereafter PAPER1;
Buote 2000d, hereafter PAPER3).

M87 is arguably the best target for study of its hot plasma with
\rosat because it is the brightest nearby galaxy, group, or cluster
that also possesses an ambient gas temperature ($\sim 2$ keV) that
lies within the bandpass of the PSPC. For these reasons M87 has been
the subject of numerous X-ray studies. The early analysis by
\citet{crc82} of the \einstein FPCS spectral data of M87 provided the
first evidence of a multiphase hot ISM in a galaxy cluster, though
this claim was disputed by \citet{tsai} using data from the other
\einstein detectors with lower energy resolution.

\citet{daw91} discovered intrinsic absorption in the \einstein SSS
data of M87 which was later confirmed with \asca \citep{adf}. In these
studies the absorber is assumed to be cold gas with solar abundances
similar to the intervening cold gas in the Milky Way.  The useful
bandpasses of the \einstein SSS and \asca SIS extend down to $\sim
0.5$ keV, and thus the \rosat PSPC whose bandpass extends down to 0.1
keV should be even more sensitive to soft X-ray absorption from cold
gas (or dust).

However, in the first detailed spectral analysis of the PSPC data of
M87 \citet{nb} mention in passing that when the hydrogen column
density is treated as a free parameter they obtain values {\it less}
than Galactic. Although not emphasized by \citet{nb} this result is
inconsistent with significant absorption from cold gas. A later \rosat
analysis by \citet{af} restricted consideration only to energies above
0.4 keV and also finds no evidence for absorption from cold gas in the
PSPC data of M87. But a subsequent re-analysis by these authors now
finds evidence for such absorption (Sanders, Fabian, \& Allen
2000). The confused status of intrinsic X-ray absorption and
multiphase gas in M87 warrants re-examination.

The possible connection between the cooling flow and AGN in M87 has
also received considerable attention recently. It has been known for
some time (e.g., B\"{o}hringer 1997) that the X-ray isophotes are
somewhat distorted within $\sim 30$ kpc of the center of M87, and
these distortions appear to correlate with the radio emission. A new
deep radio image obtained by \citet{owen} shows that the radio
emission fills the central $\sim 30$ kpc region. The spatial
correlation between radio and X-ray emission in M87 has been
interpreted as evidence that the AGN has refashioned the hot ISM and
stifled the cooling flow in the central region \citep{binney99}.

Therefore, we re-analyze the PSPC data primarily to search for oxygen
absorption and warm gas in M87 similar to that found for the lower S/N
galaxies and groups in PAPER1 and PAPER3. We also investigate cooling
flow models in an attempt to clarify the possible connection between
the AGN and cooling flow. The paper is organized in follows. In \S
\ref{obs} we present the observations and discuss the data
reduction. The deprojection procedure is summarized in \S
\ref{xdeproj}. We discuss the spectral fitting of single-phase models
in \S \ref{single} and of multiphase models in \S \ref{multi}.  In \S
\ref{disc} we discuss the results and in \S \ref{conc} we present our
conclusions.

\section{Observations and Data Preparation}
\label{obs}

\begin{table*}[t] \footnotesize
\begin{center}
\caption{\rosat PSPC Observations of M87\label{tab.obs}}
\begin{tabular}{cccccccc} \tableline\tableline\\[-7pt]
& \nh & & Date & Exposure & \multicolumn{2}{c}{Total Source Flux} & Background\\
$z$ & ($10^{20}$ cm$^{-2}$) & Sequence No & (Mo/yr) & (ks) &
(\ctspers) & (\ergcms) & (\ctspers\thinspace arcmin$^{-2}$)\\ \tableline\\[-7pt] 
0.00430 & 1.8-2.0 & rp800187n00 & 6/92 & 10.1/9.4 & 25.3 & $3.0\times
10^{-10}$ & 1.5E-3\\
&	& rp800365n00 & 7/92 & 9.9/9.5\\
&     & rp800365a01 & 12/92 & 10.0/9.7\\ \tableline \\[-35pt]
\end{tabular}
\tablecomments{Galactic Hydrogen column density (\nh) is taken from
\citet{lieu96}. The Exposure column lists first the raw exposure time
and second the final filtered exposure.  The total source count flux
is computed within a radius of $18.8\arcmin$ or 98 kpc assuming $D=18$
Mpc; i.e., within the PSPC central ring. Both the source flux and
background count rate are evaluated over 0.2-2.2 keV.}
\end{center}
\end{table*}

We obtained the \rosat PSPC data of M87 from the public data archive
maintained by the High Energy Astrophysics Science Archive Research
Center (HEASARC). The properties of the observations are listed in
Table \ref{tab.obs}. The three observations have essentially identical
exposures and pointings (i.e., centered on M87).  Each observation was
reduced separately, and we refer the reader to Buote (2000c, hereafter
PAPER2) for details of our implementation of the standard data
reduction procedures.

\subsection{Background Estimation}
\label{bkg}

Background spectra were obtained from source-free regions near the
edge of the PSPC fields; i.e., $\sim 50\arcmin$ from the field
centers.  The principal advantage of a local background estimate is
that residual contamination from solar X-rays and any other long-term
background enhancements \citep{snow2} are fully accounted for in the
ensuing spectral analysis.  The disadvantage is that the extended
X-ray emission of M87 and the Virgo cluster fills the entire PSPC
field and must be disentangled from the true background. (Note that
the background cluster emission is accounted for in the deprojection
analysis -- see \S 3.1 of PAPER2.)

We extracted the background spectra separately from each observation
and then added them together. Since each observation occurred after
the October 1991 gain change the Redistribution Matrix File (RMF),
which specifies the channel probability distribution for a photon, is
the same for each. Since, however, the detector location of background
regions are in general slightly different for the different
observations because of slight aspect differences, we averaged their
respective Auxiliary Response Files (ARFs) which contain the
information on the effective area as a function of energy and detector
location.

To obtain an estimate of the true background we follow the procedure
outlined in PAPER2 and fit the composite background spectrum with a
model after subtracting the particle background.  We represent the
cosmic X-ray background by a power law with photon index $\Gamma=1.4$
and the Galactic emission by two soft thermal components following
\citet{cfg}. (Actually, for M87 we find that only one of these soft
thermal components is required with a temperature $\approx 0.15$ keV.)
To account for emission from the extended cluster emission we include
another thermal component with variable temperature.  Each component
is modified by the Galactic hydrogen column density listed in Table
\ref{tab.obs}.

This composite model provides a good fit to the background spectrum
($\chi^2=193$ for 185 dof), and the additional thermal component
representing the cluster emission is clearly required by the data;
i.e., $\chi^2=352$ for 181 dof without the cluster component.  The
temperature of the cluster component is $1.5^{+0.4}_{-0.2}$ keV (90\%
confidence) which is somewhat less than the $\sim 3$ keV temperature
inferred by \citet{nb} over the same regions of the PSPC data. These
differences are probably the result of the different models used to
compute the background. If, for example, we do not include the cosmic
power law component in our fits then we obtain a temperature of
$\approx 2.5$ keV similar to \citet{nb}. (Note that the temperature of
the cluster component is very insensitive to the metallicity.)

Since the diffuse background is most important in the softest energy
channels the stability of the background estimate to various
assumptions can be assessed by considering the background flux between
0.2 and 0.3 keV. (The need for the 0.2 keV lower limit arises from PSF
considerations -- see below in \S \ref{src}.) Let us compare the
0.2-0.3 keV flux obtained from our best-fitting background model to a
similar model where instead the cluster component temperature is fixed
at 3 keV. We find that the 0.2-0.3 keV fluxes of the background models
(after excluding the cluster components) differ by only 14\% with the
best-fitting model having the slightly larger 0.2-0.3 keV flux. (We
also prefer the slightly larger background estimate from the
best-fitting model as it allows a more conservative assessment of any
intrinsic soft emission in M87.)  The background rate of the
best-fitting model (i.e., after excluding the 1.5 keV cluster
component) is $1.5\times 10^{-3}$ cts s$^{-1}$ arcmin$^{-2}$ (Table
\ref{obs}) which is very similar to the value of $\approx 1.0\times
10^{-3}$ cts s$^{-1}$ arcmin$^{-2}$ obtained for PSPC observations of
three other Virgo galaxies in PAPER2.

We subtract the background spectrum from each source annulus (see
below) following the procedure described in \S 4.1.4 (equation 1) in
PAPER2. That is, the background model (cosmic+Galactic) is scaled to
the source annulus accounting for differences in area, exposure,
vignetting, and detector response between the source and background
regions. The ambient Virgo emission is accounted for downstream in the
deprojection analysis discussed in \S \ref{xdeproj} (see also \S 3.1
of PAPER2).

\subsection{Source Spectra}
\label{src}

Since our interest lies principally in the central regions where any
intrinsic oxygen absorption is expected to be most pronounced, we
confine our analysis of the X-ray emission of M87 to the region within
the central ring of the PSPC (radius $20\arcmin$).  This is also the
best calibrated region of the PSPC and is not affected by shading by
the radial spokes of the PSPC support structure.  

The source spectra were extracted in concentric circular annuli
located at the X-ray centroid (computed within a $2\arcmin$ radius)
such that for each annulus the width was $\ge 1\arcmin$ and the
background-subtracted counts was larger than some value ($\approx
70000$) chosen to minimize uncertainties on the spectral parameters
for each system while maintaining as many annuli as possible (see
Figure \ref{fig.m87}.  Data with energies $\le0.2$ keV were excluded
to insure that the PSF was $<1\arcmin$ FWHM. For our on-axis sources
$\sim 99\%$ of the PSF at 0.2 keV is contained within $R=1\arcmin$
\citep{pspcpsf}.  Any background sources that were identified by
visual examination of the image were masked out before the extraction.

\section{Deprojection Method}
\label{xdeproj}

To obtain the three-dimensional properties of the X-ray emitting gas
we deproject the data assuming spherical symmetry using the
``onion-peeling'' technique pioneered by \citet{deproj}.  That is, one
begins by determining the emission in the bounding annulus and then
works inwards by subtracting off the contributions from the outer
annuli. A full account of our implementation of this procedure is
given in PAPER2. 

Since we confine our analysis to the region within the central ring of
the PSPC there exists significant emission from M87/Virgo outside of
the chosen bounding annulus. This constitutes a ``cluster background''
component that needs to be subtracted from each source annulus in
addition to the cosmic+Galactic background discussed in \S
\ref{bkg}. Following \S 3.1 of PAPER2 we represent the spatial
distribution of this cluster background by a $\beta$-model with the
same spectrum as the bounding annulus. (X-ray emissivity of a $\beta$
model is $\propto r^{-6\beta}$ at large radius.) As in PAPER2 we find
that the derived temperatures, abundances, and column densities are
very insensitive to typical values of $\beta$ for galaxies and
clusters and thus we employ $\beta=2/3$ for all results presented in
\S \ref{single} and \S \ref{multi}.

We find that the radial profiles of the temperatures, Fe abundances,
and column densities are generally well behaved. Hence, unlike for the
galaxies and groups analyzed in PAPER2 we do not perform the
quasi-regularization of the temperatures and abundances described in
\S 3.3 of PAPER2. The greater stability of parameter values can be
attributed to the much larger S/N of the M87 data.

(We mention that the deprojection procedure is invalid if there are
gaps between annuli. This is another reason why we confined our
analysis within the central ring of the PSPC.)

\vskip 1cm

\section{Single-Phase Analysis}
\label{single}

\begin{figure*}[t]
\centerline{\psfig{figure=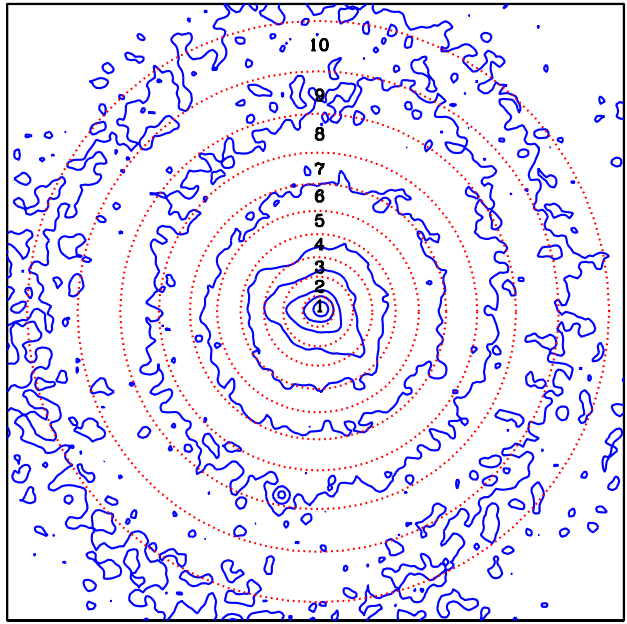,height=0.4\textheight}}
\caption{\label{fig.m87} Contour plot of the \rosat PSPC image (0.4 -
2.0 keV) of M87 corresponding to observation rp800187n00 (see Table
\ref{tab.obs}). For display only the image was binned into
$15\arcsec\times 15\arcsec$ pixels and smoothed with a gaussian filter
($\sigma=1$ pixel). Each contour is separated by a factor of 2 in
surface brightness. The dotted (red) circles demarcate the source
extraction annuli (see \S \ref{src}) defined as follows in units of
arcminutes: annulus \#1 0-1, \#2 1-2.25, \#3 2.25-3.5, \#4 3.5-5, \#5
5-6.5, \#6 6.5-8.25, \#7 8.25-10.25, \#8 10.25-12.75,\#9 12.75-15.5,
and \#10 15.5-18.75}
\end{figure*}

\begin{table*}[t] \footnotesize
\begin{center}
\caption{Quality of Spectral Fits ($\chi^2$/dof) for Single-Phase
Models\label{tab.1t}}
\begin{tabular}{ccccccccc} \tableline\tableline\\[-7pt]
Annulus & 0NH\_0Fe & 1NH\_0Fe & 0NH\_1Fe & 1NH\_1Fe & 0NH\_0Fe\_E & 1NH\_0Fe\_E & 0NH\_1Fe\_E & 1NH\_1Fe\_E\\ \tableline\\[-7pt] 
1  & 1678.0/186 & 583.8/185  & 1254.1/185 & 507.1/184   & 821.4/184 & 333.2/183 &  332.2/183 & 323.4/182\\
2  & 1056.1/189 & 784.0/188  & 1019.2/188 & 469.3/187   & 292.6/187 & 288.6/186 &  287.6/186 & 278.9/185\\
3  & 1097.2/190 & 647.1/189  & 1027.1/189 & 437.6/188   & 354.2/188 & 301.6/187 &  293.3/187 & 280.4/186\\
4  & 1248.4/190 & 617.8/189  & 1088.9/189 & 473.0/188   & 341.9/188 & 302.2/187 &  268.2/187 & 246.6/186\\
5  & 1355.6/190 & 567.7/189  & 1008.7/189 & 536.2/188   & 568.2/188 & 419.4/187 &  338.5/187 & 303.9/186\\
6  & 1358.7/190 & 424.2/189  & 909.8/189  & 397.9/188   & 509.8/188 & 307.7/187 &  242.2/187 & 222.7/186\\
7  & 1429.9/190 & 514.5/189  & 966.0/189  & 498.8/188   & 512.0/188 & 388.8/187 &  280.8/187 & 237.4/186\\
8  & 1383.7/190 & 420.4/189  & 860.0/189  & 415.4/188   & 375.3/188 & 346.8/187 &  251.0/187 & 227.6/186\\
9  & 1220.5/189 & 386.5/188  & 743.4/188  & 386.2/187   & 413.5/187 & 343.3/186 &  252.5/186 & 228.7/185\\
10 & 649.0/190  & 263.6/189  & 466.2/189  & 255.0/188   & 464.6/188 & 213.9/187 &  173.5/187 & 164.6/186
\\ \tableline \\[-35pt]
\end{tabular}
\tablecomments{Models are labeled according to whether \nh\, (standard
absorber) and/or the Fe abundance are free parameters in the fits. A
``0'' indicates the parameter is fixed and ``1'' indicates the
parameter is variable. Models with ``E'' have an intrinsic oxygen
absorption edge.}
\end{center}
\end{table*}

We initially focus our analysis on ``single-phase'' models of the hot
gas; i.e., a single value for the temperature and density describes
the emission of hot plasma at each radius. This is the simplest case
and would seem {\it a priori} to be most appropriate for the low
energy resolution of the \rosat data. Below in \S \ref{multi} we
consider multiphase models.  As in previous studies we use the \mekal
plasma code (Mewe, Gronenschild, \& van den Oord 1985; Kaastra \& Mewe
1993; Liedahl, Osterheld, \& Goldstein 1995) to represent the emission
of hot gas at a single temperature.

We take the solar abundances in \xspec to be those given by the
\citet{feld} table, because it contains the correct value of the solar
Fe abundance, $\rm Fe/H=3.24\times 10^{-5}$ by number. The popular
table of \citet{ag} used in \xspec has an old value for Fe/H which is
too large by a factor of 1.44 as initially pointed out by
\citet{im}. See \S 4.1.3 of PAPER2 for further discussion.

According to the recent deconvolution analysis of \asca data by
\citet{daw00} the Fe abundance of M87 is $\sim 0.9\solar$ within the
central arcminute, remains essentially constant at $\sim 0.6\solar$
out to $\sim 9\arcmin$, and then falls to an average value of $\sim
0.4\solar$ between $\sim 9\arcmin -19\arcmin$ (all abundances scaled
to meteoritic solar). These results agree with the multitemperature
models of the \asca SIS data obtained by fitting the integrated
spectra within $6\arcmin$ of M87 by \citet{bcf}. Consequently, when we
fix the Fe abundance in selected models we assign it the value of
$0.78\solar$ appropriate for the two-temperature model of \citet{bcf}.

Following standard practice we represent the soft X-ray absorption
arising from the Milky Way by cold material with solar abundances
distributed as a foreground screen at zero redshift. In this standard
absorption model the X-ray flux is diminished according to
$A(E)=\exp(-\nh\sigma(E))$, where \nh\, is the hydrogen column density
and $\sigma(E)$ is the energy-dependent photo-electric absorption
cross section for a cold absorber with solar abundances.

\citet{lieu96} have shown that the Galactic column density varies over
the PSPC field of M87. Inspection of their Figure 2 reveals that
\nhgal\, varies almost radially outward from M87 taking a minimum
value of $1.8\times 10^{20}$ cm$^{-2}$ near the center and a value of
$2.0\times 10^{20}$ cm$^{-2}$ near our bounding annulus (\#10). Since
intermediate values of \nhgal\, appear to prevail at intermediate
radii we set $\nhgal=1.9\times 10^{20}$ cm$^{-2}$ in those cases where
we fix \nhgal\, in our models. In other instances we allow \nh\, to be
a free parameter in our fits to indicate any excess absorption
intrinsic to a galaxy or group and also to allow for any errors in the
assumed Galactic value and for any calibration uncertainties. Note
that in this standard model \nh\, is measured as a function of
two-dimensional radius, $R$, on the sky in contrast to the emission
parameters (e.g., gas temperature, Fe abundance) that are measured as
a function of three-dimensional radius, $r$.

The photo-electric absorption cross sections used in this paper are
given by \citet{phabs}. It has been pointed out by \citet{ab} that the
He cross section at 0.15 keV is in error by 13\% in the \citet{phabs}
compilation. Recently a new version of \xspec (v11) has been released
which updates the \citet{phabs} table to include the more accurate He
cross sections of \citet{hephabs}. We find that when using these more
accurate He cross sections we obtain hydrogen column densities that
are 5\%-10\% larger than when using the original \citet{phabs} He
cross sections in \xspec v10. Where appropriate we mention these small
differences below.

\subsection{Standard Absorber}
\label{std}

\begin{figure*}[t]
\parbox{0.49\textwidth}{
\centerline{\psfig{figure=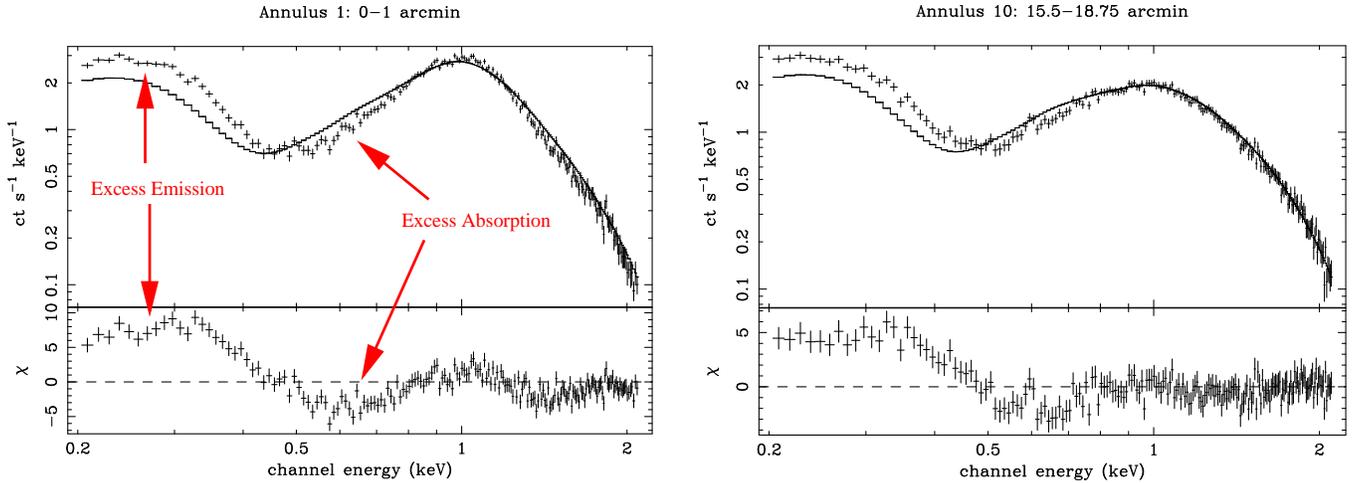,height=0.26\textheight}
}}
\parbox{0.49\textwidth}{
\centerline{\psfig{figure=1t_gnh_out.eps,angle=-90,height=0.26\textheight}
}}
\caption{\label{fig.1t_spec} \rosat PSPC spectra of the (Left) inner
annulus and the (Right) outer annulus. The fitted models have hydrogen
column density fixed to Galactic and Fe abundance fixed to
$0.78\solar$; i.e., model 0NH\_0Fe of Table \ref{tab.obs}.}
\end{figure*}

\begin{figure*}[t]
\parbox{0.32\textwidth}{
\centerline{\psfig{figure=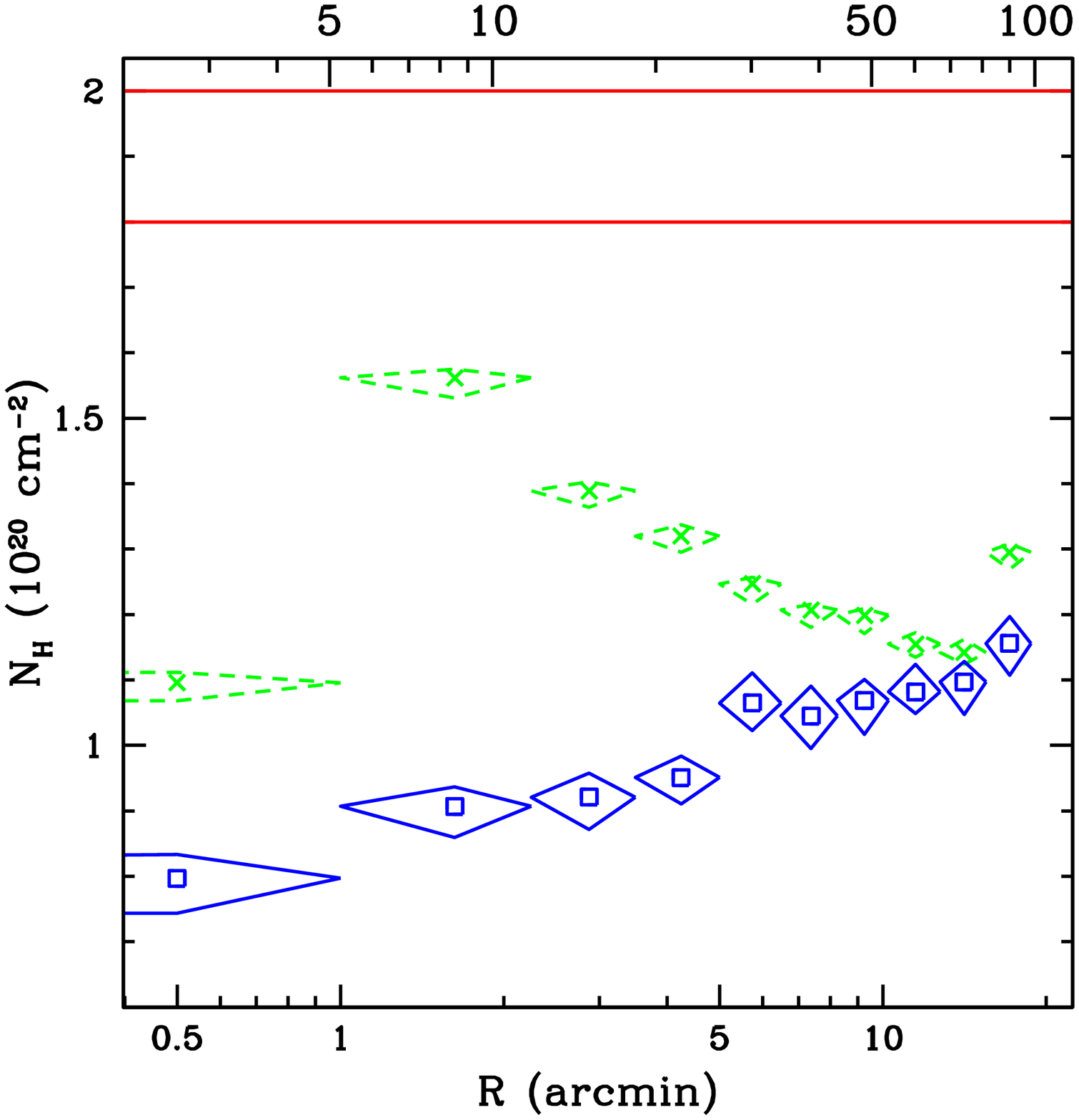,height=0.24\textheight}}
}
\parbox{0.32\textwidth}{
\centerline{\psfig{figure=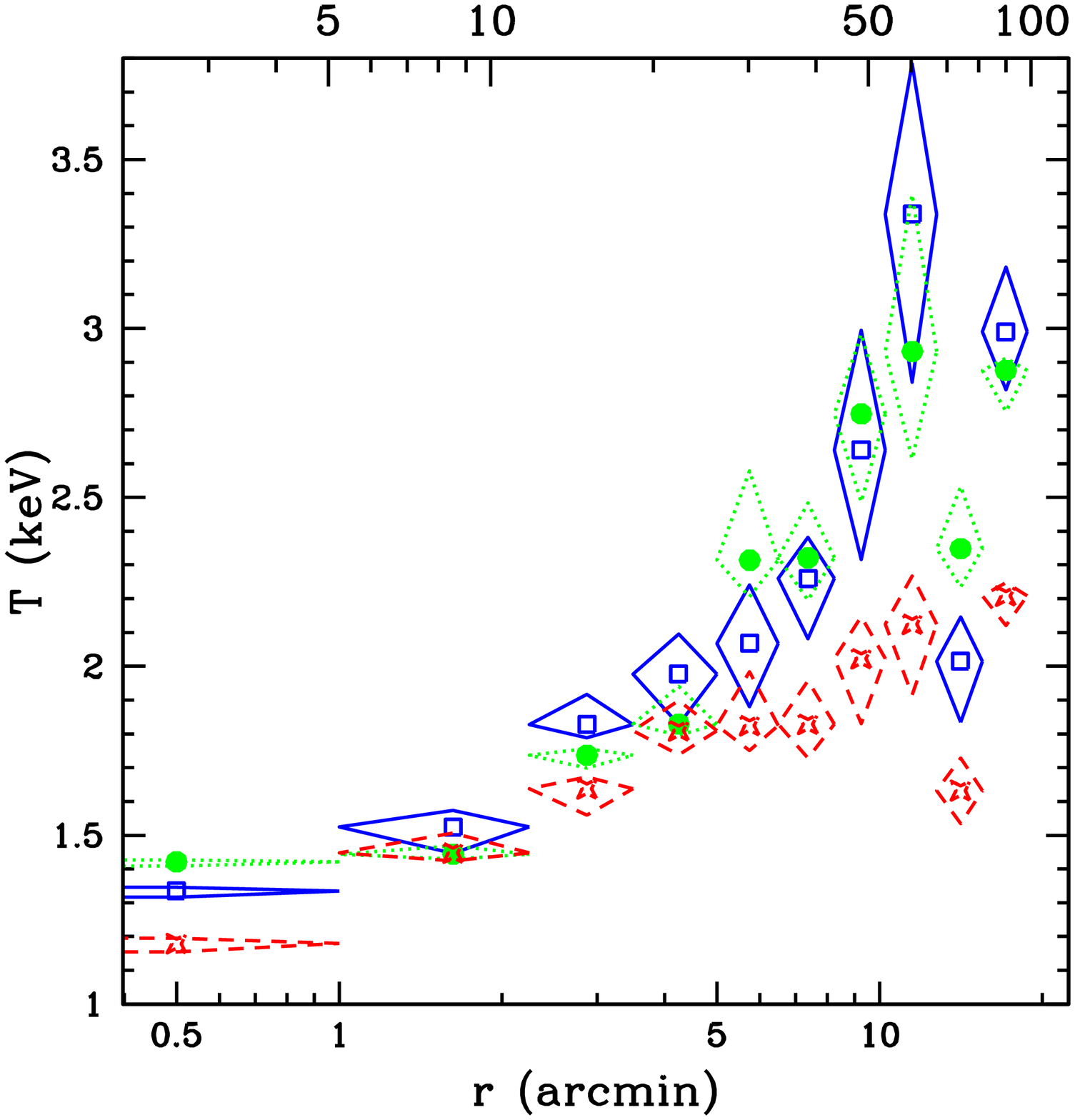,height=0.24\textheight}}
}
\parbox{0.32\textwidth}{
\centerline{\psfig{figure=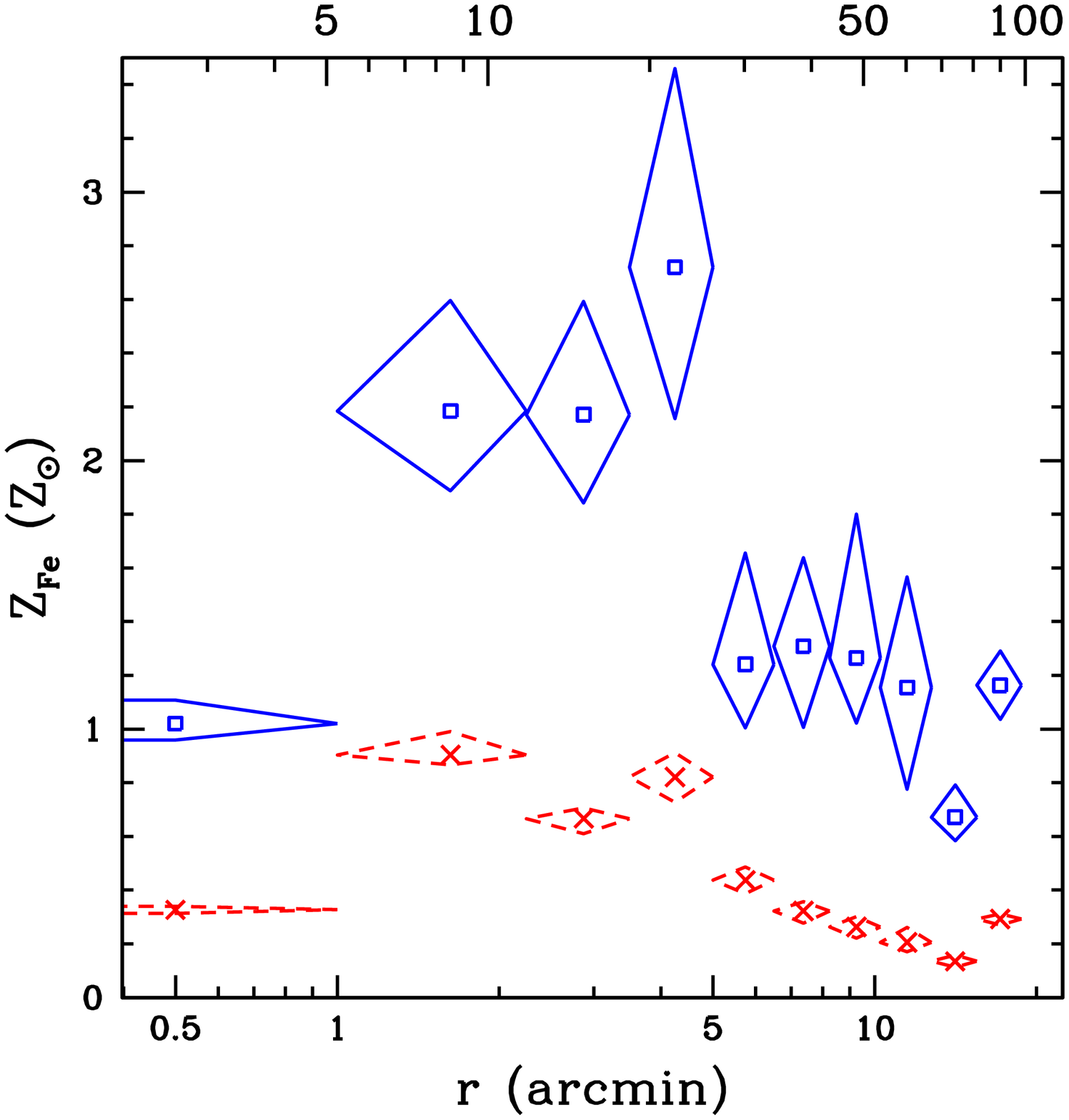,height=0.24\textheight}}
}
\caption{\label{fig.1t} Fitted model parameters for selected
single-phase models of Table \ref{tab.obs} as a function of radius
(units -- bottom: arcminutes, top: kpc): (Left) Hydrogen column
density profile of standard absorber for model 1NH\_1Fe denoted by
(blue) boxes for best-fitting values and (blue) solid diamonds for
68\% error bars. Also shown is model 1NH\_0Fe denoted by (green)
crosses and dashed diamonds. The solid lines (red) indicate the range
of Galactic column density: (Middle) Temperature profiles of models
1NH\_1Fe (boxes/solid diamonds/blue), 0NH\_1Fe (crossed/dashed
diamonds/red), and 1NH\_0Fe (circles/dotted diamonds/green): (Right)
Fe abundance profiles for models 1NH\_1Fe (boxes/solid diamonds/blue)
and 0NH\_1Fe (crossed/dashed diamonds/red).}
\end{figure*}

We begin with the model having the least number of free
parameters. This corresponds to setting $\nh=\nhgal$ for the standard
absorber and holding the Fe abundance fixed to $0.78\solar$ as
discussed above. The values of $\chi^2$ for this model (0NH\_0Fe) are
listed in Table \ref{tab.1t}, and the spectral fits for the inner and
outer annuli are displayed in Figure \ref{fig.1t_spec}. It is clear
from the large reduced $\chi^2$ values and the large residuals in the
spectral plots that this model provides a very poor fit to the \rosat
PSPC spectra of M87.

Two regions of the spectra are primarily responsible for the poor
fits. In the energy band 0.2-0.4 keV the data clearly lie above the
model prediction. Immediately one can conclude that there is no
intrinsic absorption from cold gas\footnote{These results also apply
for absorber models with partial covering factor $f=0.5$ precisely as
found for the galaxies and groups in PAPER3; see section 2.1 of PAPER1
and section 3.2 of PAPER3 for details. Since we find that the results
obtained when using absorber models with partial covering are
essentially identical to those obtained for the standard
uniform-screen absorber, we only discuss the uniform screen.}  for
which we would have expected to see a deficiency of flux below the
models with $\nh=\nhgal$.  Instead soft emission in excess of this
model is suggested.  In contrast, between 0.5-0.8 keV the data do lie
below the model consistent with intrinsic absorption. From inspection
of Figure \ref{fig.1t_spec} the absorption appears to be more
pronounced in the inner annulus.

These excess absorption and emission features are much too large to be
explained by calibration error. The most significant calibration
problem ever reported for the gain of the PSPC is by \citet{calpspc}
who showed that the early versions of the software to convert PHA
(Pulse Height Analyzer) to PI (Pulse Invariant) channels did not
account for a gain deficit within a few arcminutes distance of the
field center. This old problem only appeared in PI channels below 20
(i.e., below 0.2 keV) and is now almost entirely removed with the most
up-to-date software which we have used.  In principle such an
uncorrected gain deficit would appear as an absorption feature in our
analysis below 0.2 keV which is clearly unrelated to both the excess
emission between 0.2-0.4 keV and the absorption between 0.5-0.8 keV
observed in Figure \ref{fig.1t_spec}. Further testament to the reality
of these features is that they are each manifested over an energy
range comparable to the PSPC resolution [$\Delta E/E = 0.43 (E/0.93
\rm keV)^{-0.5}$]; i.e., in the notation of \citet{snow2}, who define
resolved PSPC energy bands, the 0.2-0.4 keV interval is essentially
the R2 band while 0.5-0.8 keV is essentially the R4 and R5 bands.

Although it might be tempting to equate the soft excess in the 0.2-0.4
keV band with an error in the background subtraction, this cannot be
the case. The emission of the diffuse background in the \rosat band
does peak at these low energies (e.g., Chen et al. 1997) and the
temperature of gas required to explain the soft excess is consistent
with the diffuse background (see below in \S \ref{2t}). However, the
background contributes only 0.6\% of the 0.2-0.4 flux within the inner
annulus, and thus only an (impossible) error of over two orders of
magnitude in the background level could account for the excess 0.2-0.4
keV flux.  Only for the bounding annulus (i.e., annulus \#10) is a
relatively small background error required (factor of $\approx 1.6$)
to explain the excess soft emission.  But we have taken care to
account for detector response differences between background and
source positions (\S \ref{bkg}).  The statistical error on the
background is $\sim 1\%$ and we estimate at most a 15\% systematic
error in the background level arising if we do not subtract the
``additional thermal component'' attributed to M87 (\S
\ref{bkg}). Hence, errors in the background subtraction cannot explain
the excess 0.2-0.4 keV emission.

Since the deviations from the 0NH\_0Fe model displayed in Figure
\ref{fig.1t_spec} must be intrinsic to M87, let us now examine whether
adjusting the available model parameters can improve the fits. If
\nh\, is allowed to vary in the fits while keeping the Fe abundance,
\fe, fixed (i.e., model 1NH\_0Fe in Table \ref{tab.1t}) then we obtain
large improvements in the fits in all annuli; the least amount of
improvement is observed for annuli \#2-3. Although allowing \nh\, to
vary removes a large portion of the deviations over 0.2-0.4 keV the
absorption feature over 0.5-0.8 keV remains essentially unaffected.

The fitted values of \nh(R) (see Figure \ref{fig.1t}) are $\approx
0.6\nhgal$ in most annuli with a maximum of $\approx 0.8\nhgal$ in
annulus \#2. These values are still very significantly less than the
Galactic value when considering the 5\%-10\% underestimates arising
from inaccuracies in the He cross sections (see immediately before \S
\ref{std}). These sub-Galactic columns are also unaffected by possible
errors in the plasma emission code (\mekal). Since the fitted
temperatures are always above 1 keV inaccuracies in the plasma code
near energies 0.2-0.4 keV are negligible because at these temperatures
the continuum dominates any neglected line emission which is the
principal source of error (e.g., Liedahl 1999).

If instead \fe is allowed to vary but \nh\, is held fixed at \nhgal\,
(i.e., model 0NH\_1Fe in Table \ref{tab.1t}), then the fits are also
improved though not quite so much as when \nh\, is allowed to
vary. The improvement in the fits is again confined to the lower
energies (0.2-0.4 keV) leaving the absorption deviations over 0.5-0.8
keV uncorrected. The sub-solar values of $\fe(r)$ (Figure
\ref{fig.1t}) are consistent with those obtained from the previous
\asca studies of \citet{daw00} and \citet{bcf}. The \asca measurements
of $\fe$ should be more reliable since they are determined primarily
by the 6.5 keV Fe K$\alpha$ line complex in contrast to the \rosat
determinations using the $\sim 1$ keV Fe L lines.

The temperature measurements from \asca and \sax are also more
reliable than those from \rosat because of the much better energy
resolution and the larger bandwidth. In Figure \ref{fig.1t} one
notices that for $r>5\arcmin$ the model 0NH\_1Fe has $T<2.5$ keV which
is very inconsistent with the temperatures obtained from \asca
\citep{daw00} and \sax \citep{dacri}. On the other hand, if both \nh\,
and \fe are varied (i.e., model 1NH\_1Fe in Table \ref{tab.1t}) then
$T(r)$ is consistent with \asca and \sax but now \fe is too large at
all radii (Figure \ref{fig.1t}). This latter model also has fitted
column densities that deviate from the Galactic value even more
significantly than the 1NH\_0Fe model discussed above. Interestingly,
when both \nh\, and \fe are varied significant improvement in the
fits is also observed over 0.5-0.8 keV unlike when only one of these
parameters is varied (though the overall fit quality is still poor in
all annuli except \#10). 

Hence, the single-phase model with a standard absorber is demonstrably
a very poor fit to the \rosat data. Although improvements in the fits
can be achieved by varying \nh\, and \fe the resulting fits are still
poor, the column densities are significantly below the Galactic value,
and the fitted temperatures and abundances are not each consistent
with the \asca (and \sax) data for any particular model. We are thus
compelled to conclude that a single-phase gas modified by a standard
Galactic absorber cannot describe the X-ray data of M87.

\subsection{Oxygen Edge}
\label{edge}

\begin{figure*}[t]
\centerline{\psfig{figure=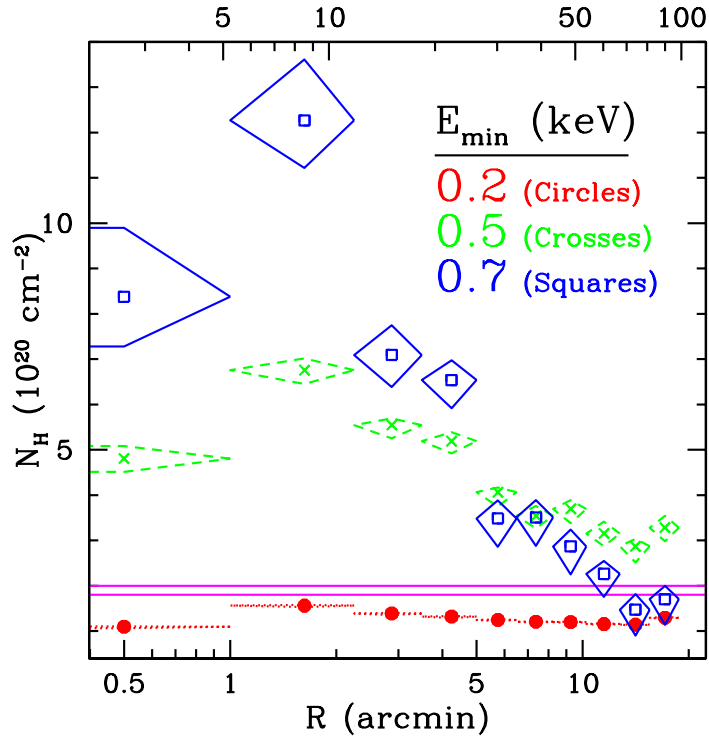,height=0.4\textheight}}
\caption{\label{fig.nhvsemin} Radial profiles of hydrogen column
density of the standard absorber for different values of the minimum
energy of the bandpass (\emin) corresponding to the single-phase model
1NH\_0Fe using the notation of Table \ref{tab.1t}. Only data with
energies $E>\emin$ are included in the fits. The solid lines (magenta)
indicate the range of Galactic column density.}
\end{figure*}

\begin{figure*}[t]
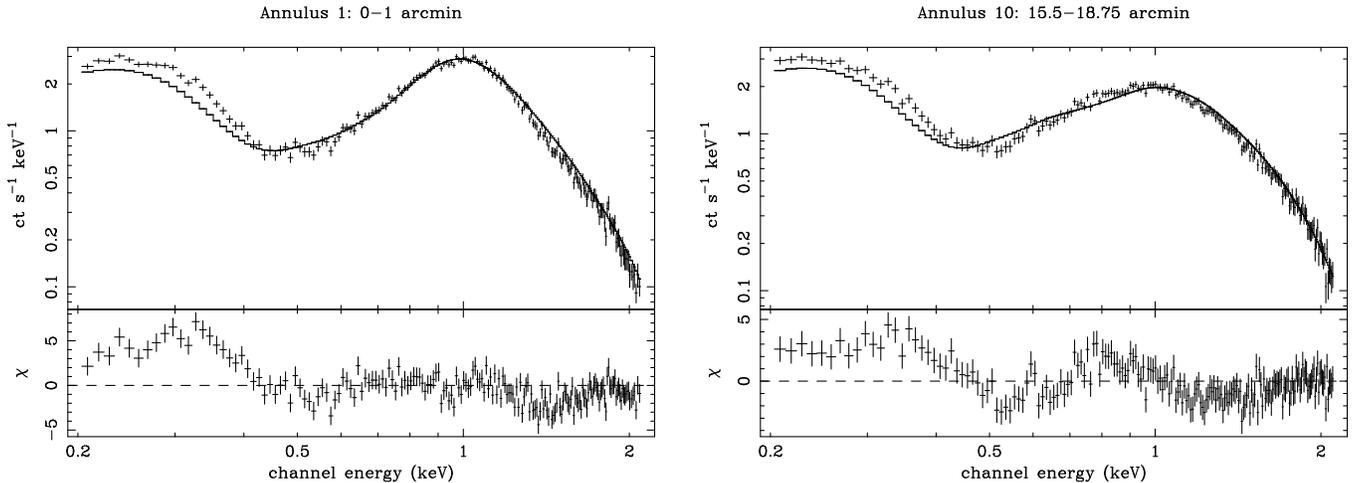

\parbox{0.49\textwidth}{
\centerline{\psfig{figure=1t_gnh_edge_in.eps,angle=-90,height=0.26\textheight}}}
\parbox{0.49\textwidth}{
\centerline{\psfig{figure=1t_gnh_edge_out.eps,angle=-90,height=0.26\textheight}}}
\caption{\label{fig.1t_edge_spec} \rosat PSPC spectra plotted as in
Figure \ref{fig.1t_spec} except that now the model includes an
intrinsic (oxygen) absorption edge; i.e., model 0NH\_0Fe\_E in Table
\ref{tab.1t}.}
\end{figure*}

\begin{figure*}[t]
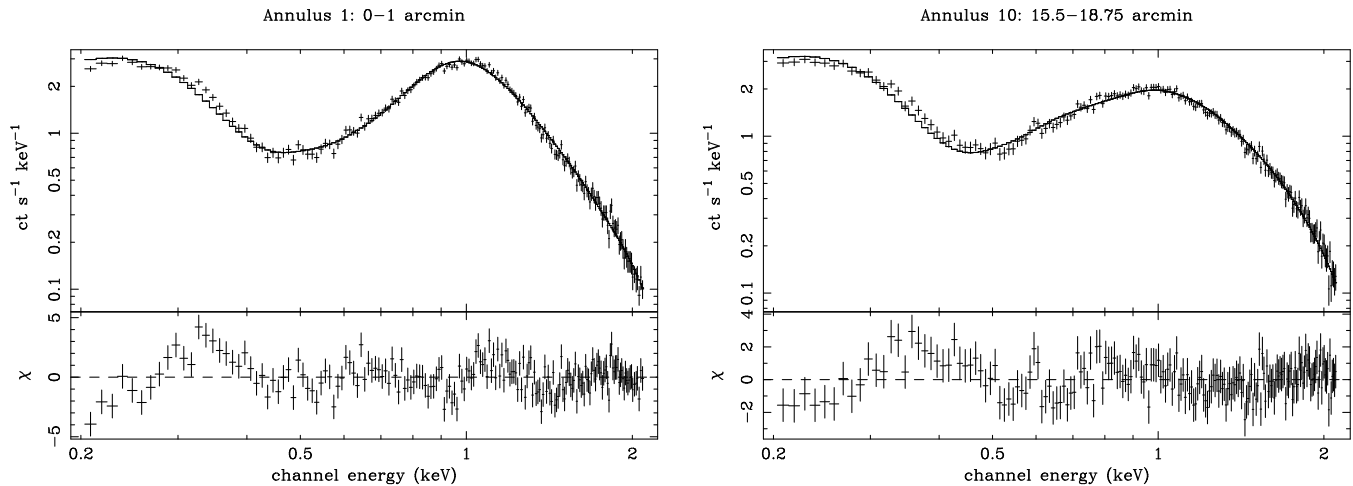

\parbox{0.49\textwidth}{
\centerline{\psfig{figure=1t_fnh_edge_in.eps,angle=-90,height=0.26\textheight}}}
\parbox{0.49\textwidth}{
\centerline{\psfig{figure=1t_fnh_edge_out.eps,angle=-90,height=0.26\textheight}}}
\caption{\label{fig.1t_edge_spec_nh} As Figure \ref{fig.1t_edge_spec}
except that now the model (with intrinsic absorption edge) also has
variable hydrogen column density for the standard absorber; i.e.,
model 1NH\_0Fe\_E in Table \ref{tab.1t}.}
\end{figure*}

\begin{figure*}[t]
\parbox{0.32\textwidth}{
\centerline{\psfig{figure=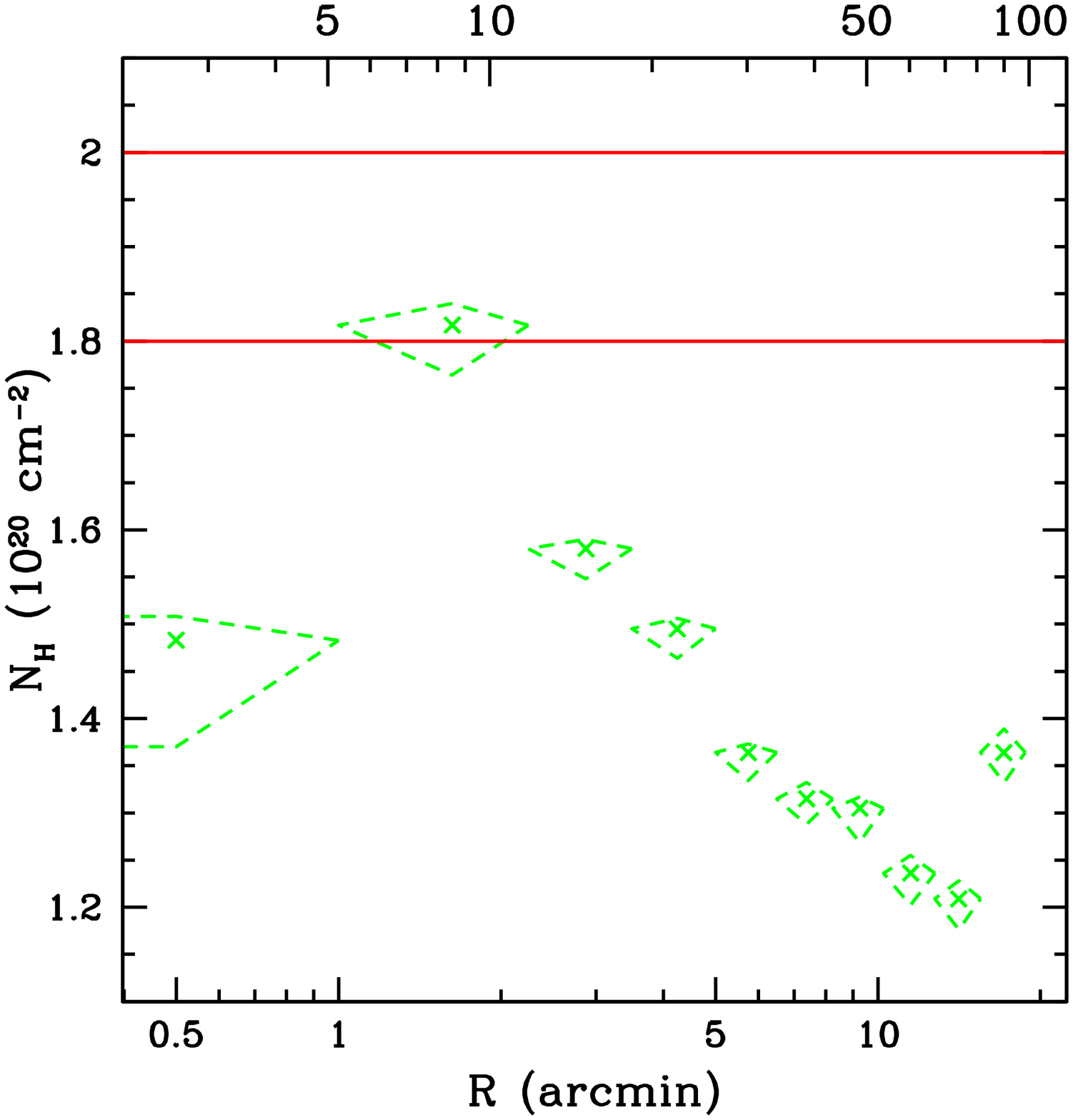,height=0.24\textheight}}
}
\parbox{0.32\textwidth}{
\centerline{\psfig{figure=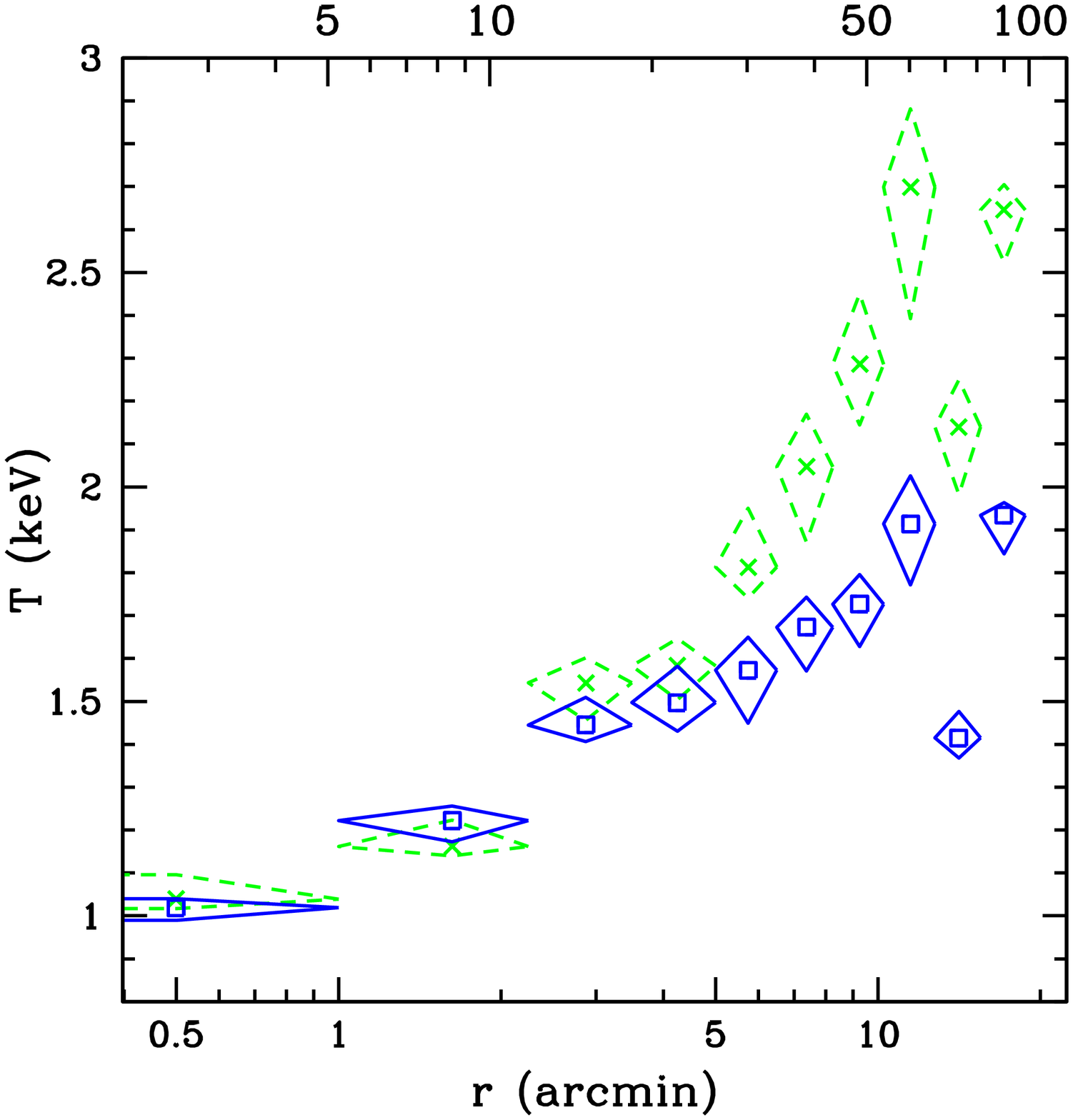,height=0.24\textheight}}
}
\parbox{0.32\textwidth}{
\centerline{\psfig{figure=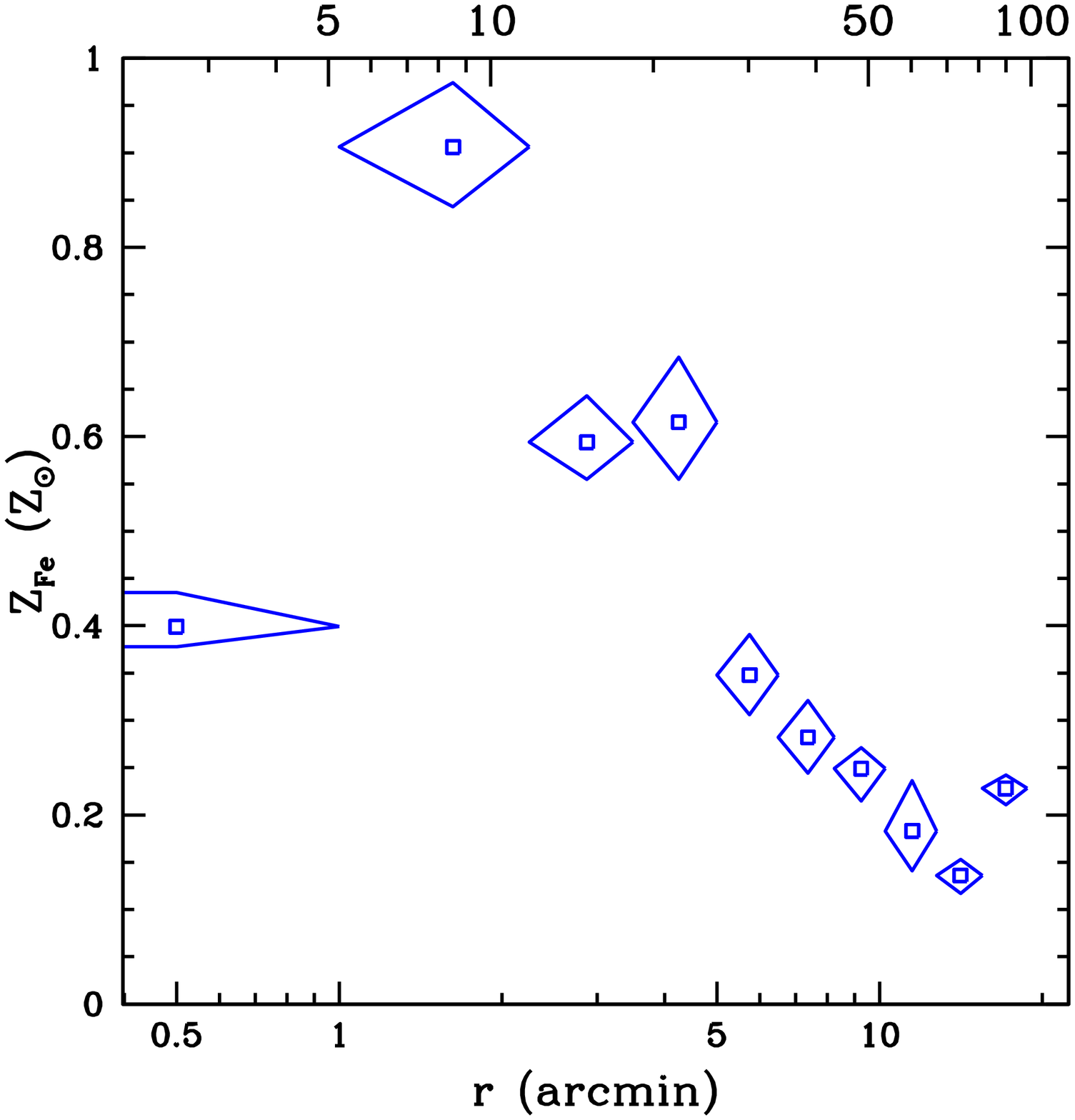,height=0.24\textheight}}
}
\caption{\label{fig.1t_edge} As Figure \ref{fig.1t} except models now
include an intrinsic oxygen edge: (Left) \nh(R) for model 1NH\_0Fe\_E;
(Middle) $T(r)$ for models 0NH\_1Fe\_E (crosses/dashed diamonds/green)
and 0NH\_1Fe\_E (boxes/solid diamonds/blue); (Right) $\fe(r)$ for
model 0NH\_1Fe\_E.}
\end{figure*}

\begin{figure*}[t]
\parbox{0.49\textwidth}{
\centerline{\psfig{figure=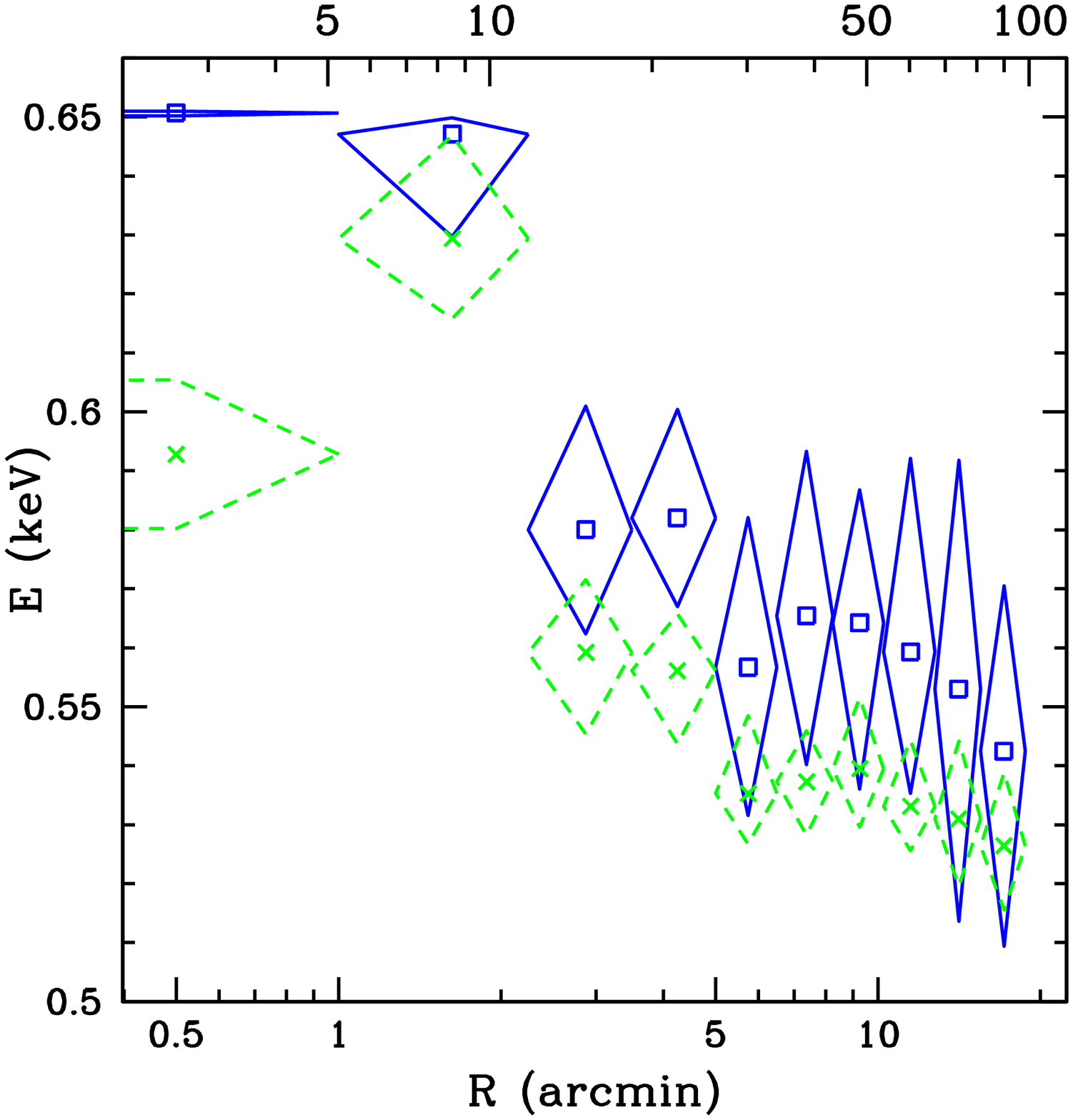,height=0.33\textheight}
}}
\parbox{0.49\textwidth}{
\centerline{\psfig{figure=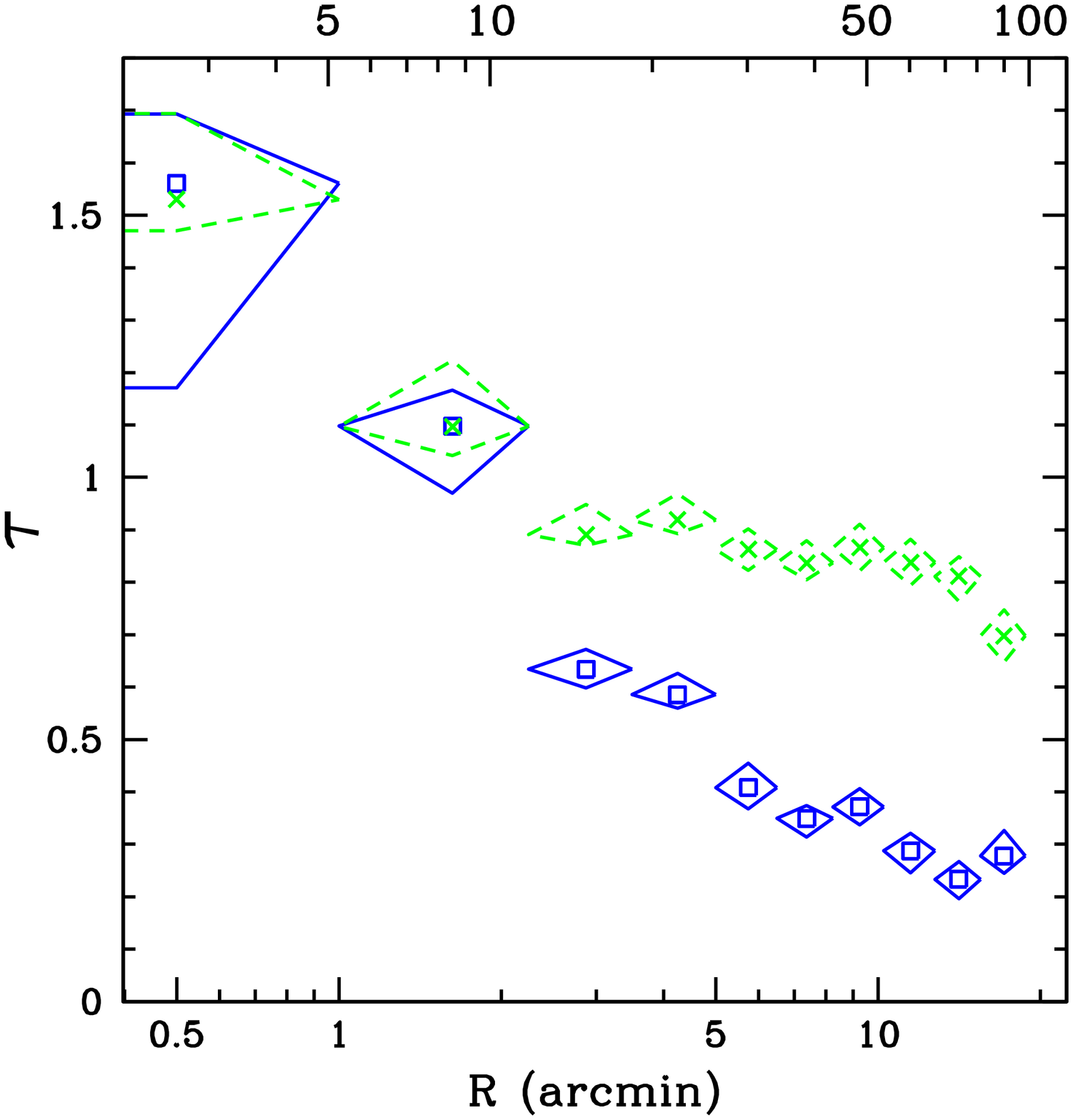,height=0.33\textheight}
}}
\caption{\label{fig.1t_edge_tau} Intrinsic oxygen edge parameters for
single-phase models: (Left) Oxygen edge energy, $E_0(R)$, for models
1NH\_0Fe\_E (crosses/dashed diamonds/green) and 0NH\_1Fe\_E
(boxes/solid diamonds/blue); (Right) Oxygen edge optical depth,
$\tau(R)$, for the same models.}
\end{figure*}

The striking residuals of excess absorption and emission apparent from
visual inspection of the spectral fits of M87 (Figure
\ref{fig.1t_spec}) provide strong support for the analogous (but less
obvious) properties deduced from the lower S/N data of the individual
elliptical galaxies and galaxy groups studied in PAPER1 and PAPER3. An
immediate result of critical importance is that since the data for
energies below 0.5 keV do not lie below the 0NH\_0Fe model in Figure
\ref{fig.1t_spec}, the low-energy PSPC data unequivocally rule out any
significant excess absorption from cold absorbing matter which has
frequently been advocated in the past for cooling flows. Instead, the
absorption feature between 0.5-0.8 keV must arise from warm material
as we have discussed extensively in PAPER1 and PAPER3.

Because of the low energy resolution of \rosat it is instructive first
to analyze the hydrogen column density obtained for the standard cold
absorber as a function of the chosen lower energy limit (\emin) of the
bandpass (see \S 3.3 of PAPER3). That is, we fitted the model 1NH\_0Fe
(defined above) to the PSPC spectrum including only data having
energies between $\emin$ and 2.2 keV. The results are displayed in
Figure \ref{fig.nhvsemin}. When $\emin\approx 0.2-0.3$ keV the fitted
column density is always similar to or less than \nhgal. As \emin\,
increases to $\sim 0.5$ keV we find that $\nh\sim 1.5\nhgal$ for $R\ga
5\arcmin$ and increases to $\nh\sim 3\nhgal$ for $R< 5\arcmin$. This
radially decreasing profile of column density becomes even more
pronounced for larger values of \emin; e.g., as shown in Figure
\ref{fig.nhvsemin} for $\emin=0.7$ keV the outer annuli have
$\nh\approx\nhgal$ whereas at smaller radii $\nh=(3-6)\nhgal$. This
trend continues up to $\emin\approx 0.8$ keV after which $\nh(R;\emin)
\approx \nh(R;0.8\, \rm keV)$ for $\emin > 0.8$ keV within the
increasingly larger uncertainties.

In order to quantify the ``warm'' absorption and yet simultaneously
fit the entire \rosat spectrum we follow our previous investigations
of galaxies and groups (PAPER1 and PAPER3) and model the intrinsic
absorption with a single absorption edge represented as, $A(E) =
\exp[-\tau(E/E_0)^{-3}]$ for $E\ge E_0$ where $E_0$ is the rest-frame
energy of the edge and $\tau$ is the optical depth. Similar to the
standard absorber we place the edge immediately in front of the source
so $\tau$ is measured as a function of 2D radius $R$. A rigorous
treatment of the warm absorber would (at least) include a series of
absorption edges from different ionization states of principally
oxygen but also of carbon and nitrogen similar to the models described
by \citet{kk} that are currently not supported in \xspec.  Our simple
parameterization of a warm absorber as a single edge is intended
primarily as a phenomenological tool to establish the existence and to
study the gross properties of the absorber which is appropriate for
the low resolution data afforded by the PSPC.

We place the edge initially at $E_0=0.532$ keV appropriate for
\ion{O}{1} but always allow $E_0$ to vary; nevertheless we shall often
refer to this as an ``oxygen edge''. The values of $\chi^2$ for models
with an oxygen edge are listed in Table \ref{tab.1t}, and in Figure
\ref{fig.1t_edge_spec} we plot the \rosat spectra for annuli \#1 and
\#10 in analogy to Figure \ref{fig.1t_spec} with the only difference
being that the oxygen edge is included in the best-fitting models;
i.e., model 0NH\_0Fe\_E.

Although the edge greatly improves the fits of all annuli, the
character of the improvement varies with radius. As seen in Figure
\ref{fig.1t_edge_spec} for annulus \#1 the oxygen edge removes most of
the residuals over 0.5-0.8 keV while leaving other deviations
(primarily at lower energies) mostly unchanged. This qualitative
behavior also applies to the next few annuli (\#2-4).  In contrast, as
seen for annulus \#10 in Figure \ref{fig.1t_edge_spec} the single
oxygen edge itself is insufficient to remove the residuals over
0.5-0.8 keV and instead allows only a relatively crude improvement by
cutting essentially midway through the 0.5-0.8 keV region; i.e., the
model is still clearly the wrong shape in the absorbing region. This
qualitative behavior applies to the outer annuli (\#5-10).

When either \nh\, of the standard absorber or \fe is allowed to vary
the fits are improved substantially in most annuli; i.e., models
1NH\_0Fe\_E and 0NH\_1Fe\_E in Table \ref{tab.1t}.  For annuli \#1-4
the improvement offered by variable \nh\, and \fe are very similar,
and as seen in Figure \ref{fig.1t_edge_spec_nh} (using model
1NH\_0Fe\_E) most of the large residuals displayed by the original
single-phase standard absorber model (Figure \ref{fig.1t_spec}) have
been removed.

The radial profiles of \nh, $T$, and \fe for these edge models are
displayed in Figure \ref{fig.1t_edge}. For annuli \#1-4 the
temperatures are consistent for both models and the Fe abundance
profile (for 0NH\_1Fe\_E) is essentially consistent with that obtained
from the \asca data \citep{daw00} (scaled to meteoritic solar Fe
abundance). The column density profile for the standard absorber is
$\sim 20\%$ below the Galactic value for annuli \#1,3-4.  This
discrepancy, though still statistically significant, is reduced to
10\%-15\% when using the more accurate He cross sections of
\citep{hephabs} in \xspec v11 (see immediately before \S \ref{std}).
Note that if both \nh\, and \fe are varied (i.e., model 1NH\_1Fe\_E)
the fitted parameters for annuli \#1-4 are consistent with those above
(though with larger errors) and the fits are slightly improved further
(Table \ref{tab.1t}).

For the outer annuli (\#5-10) although varying \nh\, improves the fits
noticeably, a much larger improvement is obtained by varying \fe.  As
can be seen in Figure \ref{fig.1t_edge_spec_nh} the variable \nh\,
model (1NH\_0Fe\_E) fits the \rosat spectrum for the bounding annulus
(\#10) quite well; the variable-\fe model (0NH\_1Fe\_E) fits even
better (Table \ref{tab.1t}). For annuli \#5-9 the variable-\fe model
fits similarly to that shown for \#10 except that the residuals below
0.5 keV are slightly more pronounced. If both \nh\, and \fe are varied
then the fits in these outer annuli are essentially formally
acceptable (Table \ref{tab.1t}).

However, the fitted parameters obtained in these outer annuli indicate
a problem with the simple single-phase, single-edge models. If \fe is
varied and \nh\, is held fixed then the inferred \fe(R) (Figure
\ref{fig.1t_edge}) is approximately consistent with the sub-solar
values obtained by \citet{daw00} for $R\ga 5\arcmin$ with \asca. But
then the temperatures of this model are always below 2 keV at large
$R$ which are very inconsistent with the 2.5-3 keV temperatures
inferred from \asca and \sax (see discussion in \S \ref{std}). On the
other hand, if \nh\, is varied and \fe is held fixed the fitted
temperatures are consistent with \asca/\sax but now $\nh\approx
(0.6-0.7)\nhgal$ at large radius.

The lower temperatures inferred for 0NH\_1Fe\_E and the sub-Galactic
columns inferred for 1NH\_0Fe\_E both imply the existence of an extra
emission component having a temperature significantly less than the
hot phase for annuli \#5-10. For the inner radii \#1-4 the
temperatures are also consistently 30\%-50\% below the \asca/\sax
values (White 2000; D'Acri et al. 1998) and (except annulus \#2) have
sub-Galactic column densities which also suggest extra emission from
cooler gas.

This cooler emission component is very likely related to the
absorption component that we have modeled with the single oxygen
edge. In Figure \ref{fig.1t_edge_tau} we plot $E_0(R)$ and $\tau(R)$
for the models shown in Figure \ref{fig.1t_edge}. For $R>5\arcmin$,
$E_0\sim 0.55$ keV and then increases to $E_0\sim 0.6-0.65$ keV at
smaller $R$. Assuming collisionally ionized gas these edge energies
imply a gas temperature ranging from $10^5-10^6$ K (e.g., Krolik \&
Kallman 1984; Sutherland \& Dopita 1993) which apparently increases
with decreasing $R$. The optical depth profile also falls with
increasing $R$ taking values near 1 for $R\la 3\arcmin$ and flattens
out for 1NH\_0Fe\_E or decreases steadily to values $\sim 0.3$ for
0NH\_1Fe\_E. Although $\tau(R)$ is quite similar to those obtained for
the galaxies and groups studied in PAPER1 and PAPER3, the values of
$E_0\sim 0.6-0.65$ keV at small radii are significantly larger and
probably indicate that the warm gas is ``warmer'' in M87.

Finally, we mention that attributing the 0.5-0.8 keV absorption
feature instead to a sub-solar O/Fe ratio of the hot gas always
results in sub-Galactic column densities for the standard absorber
similar to that shown in Figure \ref{fig.1t_spec} for model 1NH\_0Fe;
i.e., sub-solar O/Fe still implies excess soft emission. And if this
soft emission originates from ionized gas it should produce absorption
near 0.5 keV. Moreover, sub-solar O/Fe models never fit as well as the
oxygen-edge models except when \fe is held fixed and \nh\, is
varied. In this case the sub-solar O/Fe models fit comparably to the
oxygen-edge models at larger radius (annuli \#5-10) but at smaller
radii still have $\chi^2$ values that are larger by 50-100. It should
also be noted that these sub-solar O/Fe models all result in oxygen
abundances that are zero everywhere which is in serious conflict with
the theoretical expectation of approximately solar abundance ratios
(e.g., Renzini et al. 1993). If O/Fe is kept to a minimum of $\sim
1/2$ solar then no sub-solar O/Fe model fits as well as a model with
an oxygen edge. Hence, a sub-solar value for O/Fe of the hot gas
cannot reasonably describe either the excess absorption over 0.5-0.8
keV or the excess soft emission over 0.2-0.4 keV.

(We also note that a very sub-solar value of O/Fe is inconsistent with
the modest super-solar Si/Fe and S/Fe ratios measured by \asca (e.g,
Buote et al 1999); i.e., the $\alpha$/Fe ratios should all give a
consistent picture of the supernova enrichment.)

\section{Multiphase Analysis}
\label{multi}

\begin{table*}[t] \footnotesize
\begin{center}
\caption{Quality of Spectral Fits ($\chi^2$/dof) for Multiphase
Models\label{tab.multi}}
\begin{tabular}{ccccc|cccc} \tableline\tableline\\[-7pt]
    & \multicolumn{4}{c}{Cooling Flow} & \multicolumn{4}{c}{Two Phases}\\ 
Annulus & 0NH\_0Fe\_E & 1NH\_0Fe\_E & 0NH\_1Fe\_E & 1NH\_1Fe\_E & 0NH\_0Fe & 0NH\_1Fe & 0NH\_0Fe\_E & 0NH\_1Fe\_E  \\ \tableline\\[-7pt] 
1  & 329.2/183 & 324.9/182  & 325.7/182 & 320.1/181 & 529.8/184   & 389.3/183   & 288.1/182 & 256.2/181 \\
2  & 299.0/186 & 295.0/185  & 289.0/185 & 276.5/184 & 746.5/188   & 370.3/186   & 288.9/185 & 229.4/184 \\
3  & 312.0/187 & 308.2/186  & 295.7/186 & 275.7/185 & 597.0/188   & 375.5/187   & 283.7/186 & 244.0/185 \\
4  & 290.9/187 & 294.0/186  & 270.6/186 & 238.7/185 & 564.3/188   & 404.5/187   & 261.0/186 & 236.9/185 \\
5  & 424.1/187 & 372.2/186  & 335.0/186 & 298.2/185 & 527.0/188   & 489.0/187   & 321.1/186 & 295.8/185 \\
6  & 365.4/187 & 286.3/186  & 247.5/186 & 230.4/185 & 384.0/188   & 353.2/187   & 240.1/186 & 227.6/185 \\
7  & 422.6/187 & 330.6/186  & 267.6/186 & 232.5/185 & 473.6/188   & 452.9/187   & 257.3/186 & 242.0/185 \\
8  & 415.3/187 & 299.2/186  & 242.3/186 & 223.2/185 & 384.9/188   & 377.9/187   & 231.8/186 & 215.3/185 \\
9  & 413.0/186 & 303.3/185  & 247.9/185 & 226.6/184 & 360.5/187   & 359.1/186   & 240.5/185 & 225.3/184 \\
10 & 259.8/187 & 180.7/186  & 160.4/186 & 156.7/185 & 248.4/188   & 241.7/187   & 203.7/186 & 150.3/185 
\\ \tableline \\[-35pt]
\end{tabular}
\tablecomments{The notation is as described in the notes to Table
\ref{tab.1t}.}
\end{center}
\end{table*}

\begin{figure*}[t]
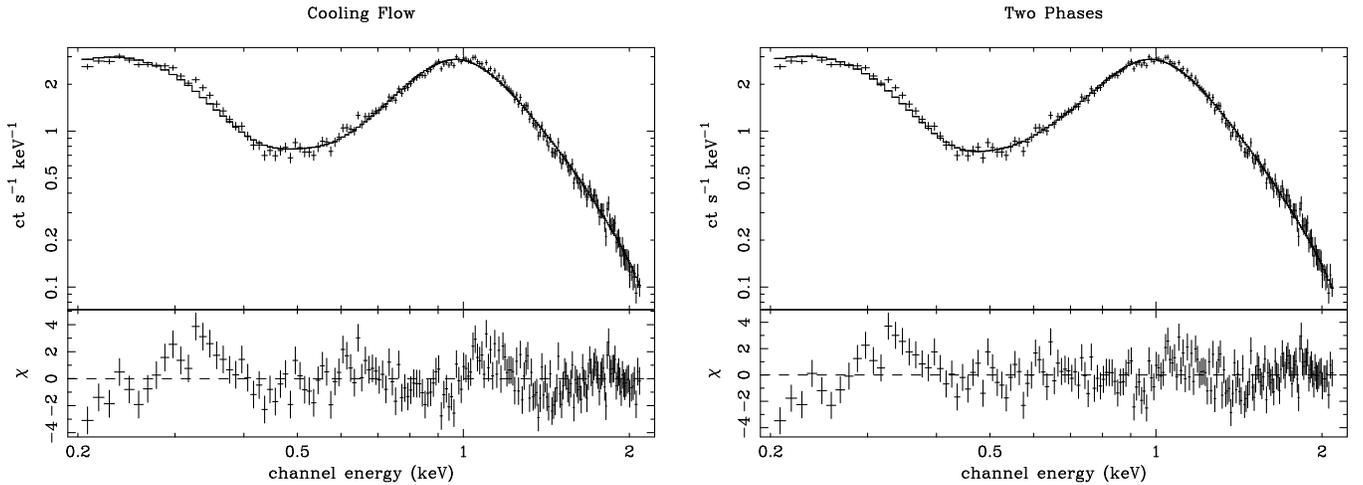

\parbox{0.49\textwidth}{
\centerline{\psfig{figure=cpcool_gnh_edge.eps,angle=-90,height=0.26\textheight}}}
\parbox{0.49\textwidth}{
\centerline{\psfig{figure=2t_gnh_edge.eps,angle=-90,height=0.26\textheight}}}
\caption{\label{fig.multi_spec} \rosat PSPC spectrum of annulus \#1
along with best-fitting models (0NH\_0Fe\_E) for (Left) a cooling flow
and (Right) a two-phase medium.}
\end{figure*}

We have seen that an attempt to describe the \rosat X-ray emission of
M87 by a single-phase hot medium fails because the intrinsic
absorption feature over 0.5-0.8 keV and the sub-Galactic hydrogen
column densities (i.e., excess soft 0.2-0.4 keV emission) imply the
existence of additional gas phases and cannot be explained by dust (\S
5.1 of PAPER3). In principle we would like to fit a general emission
model in order to determine the differential emission measure function
which fully characterizes the temperature structure of the
plasma. Unfortunately, with detectors of low (\rosat) and moderate
(\asca) energy resolution it is impossible to distinguish between
different multiphase models. \citet{bcf} showed in particular that
two-phase models can easily mimic a multiphase cooling flow spectrum
of galaxies and clusters. And in section 5.3 of \citet{b99} we showed
that even with the higher resolution of \asca the following multiphase
models of galaxies all produced similar spectra: cooling flow,
two-phase, and integrated single-phase temperature gradient; similar
results were obtained for groups in \citet{b00a}. (However, these very
different multiphase models are each easily distinguished from
isothermal models.)

Since the limited energy resolution of the PSPC prevents a detailed
exploration of different multiphase models our primary objective in
this section is to quantify the emission of the warm gas at each
radius. As in our previous studies we examine the multiphase
hypothesis in terms of the simple cases of a constant-pressure cooling
flow and a two-phase medium.  These models are instructive since they
have been widely used to interpret the X-ray spectra of galaxies and
clusters.

\subsection{Cooling Flow}
\label{cf}

\begin{figure*}[t]
\parbox{0.32\textwidth}{
\centerline{\psfig{figure=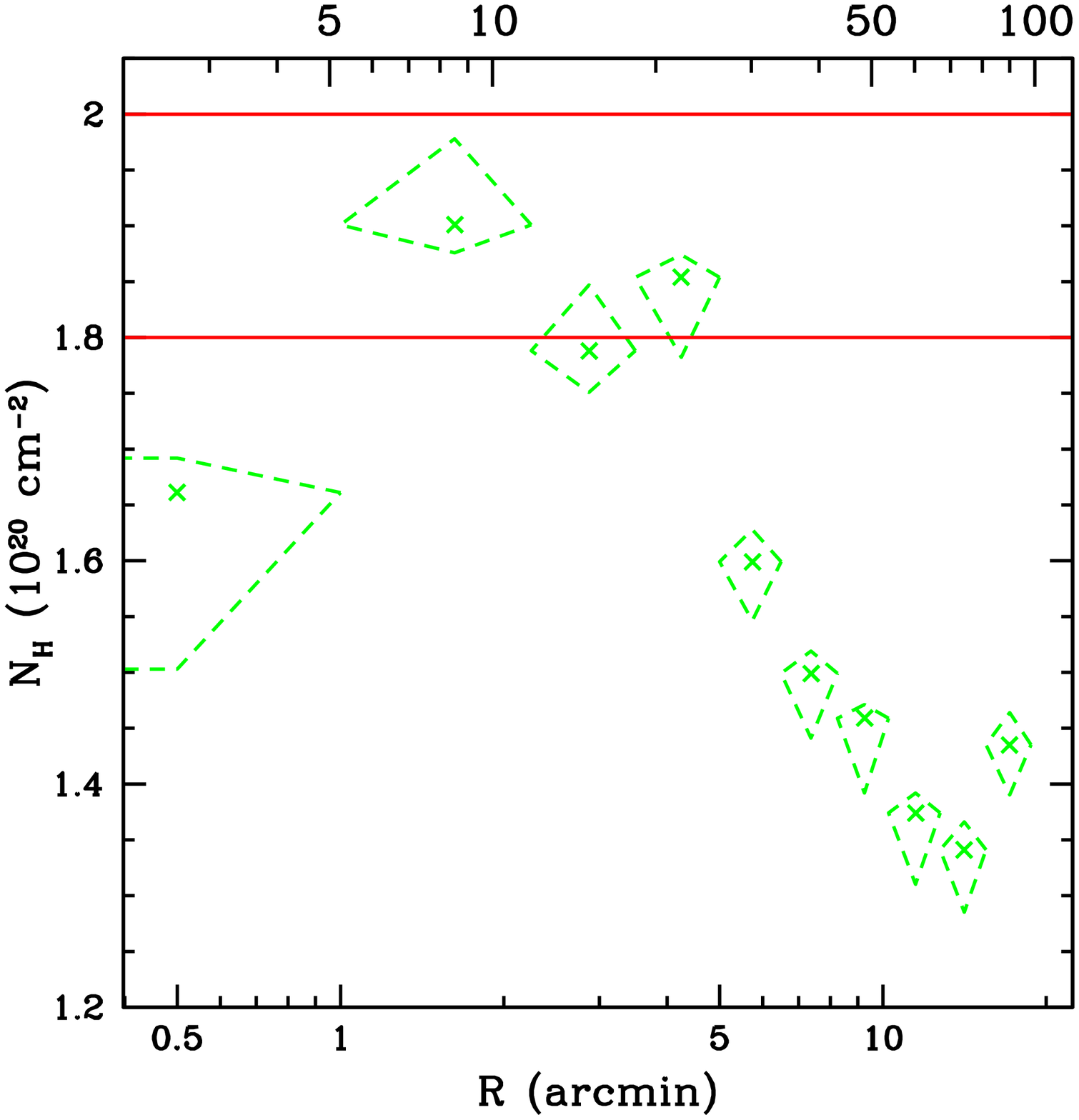,height=0.24\textheight}}
}
\parbox{0.32\textwidth}{
\centerline{\psfig{figure=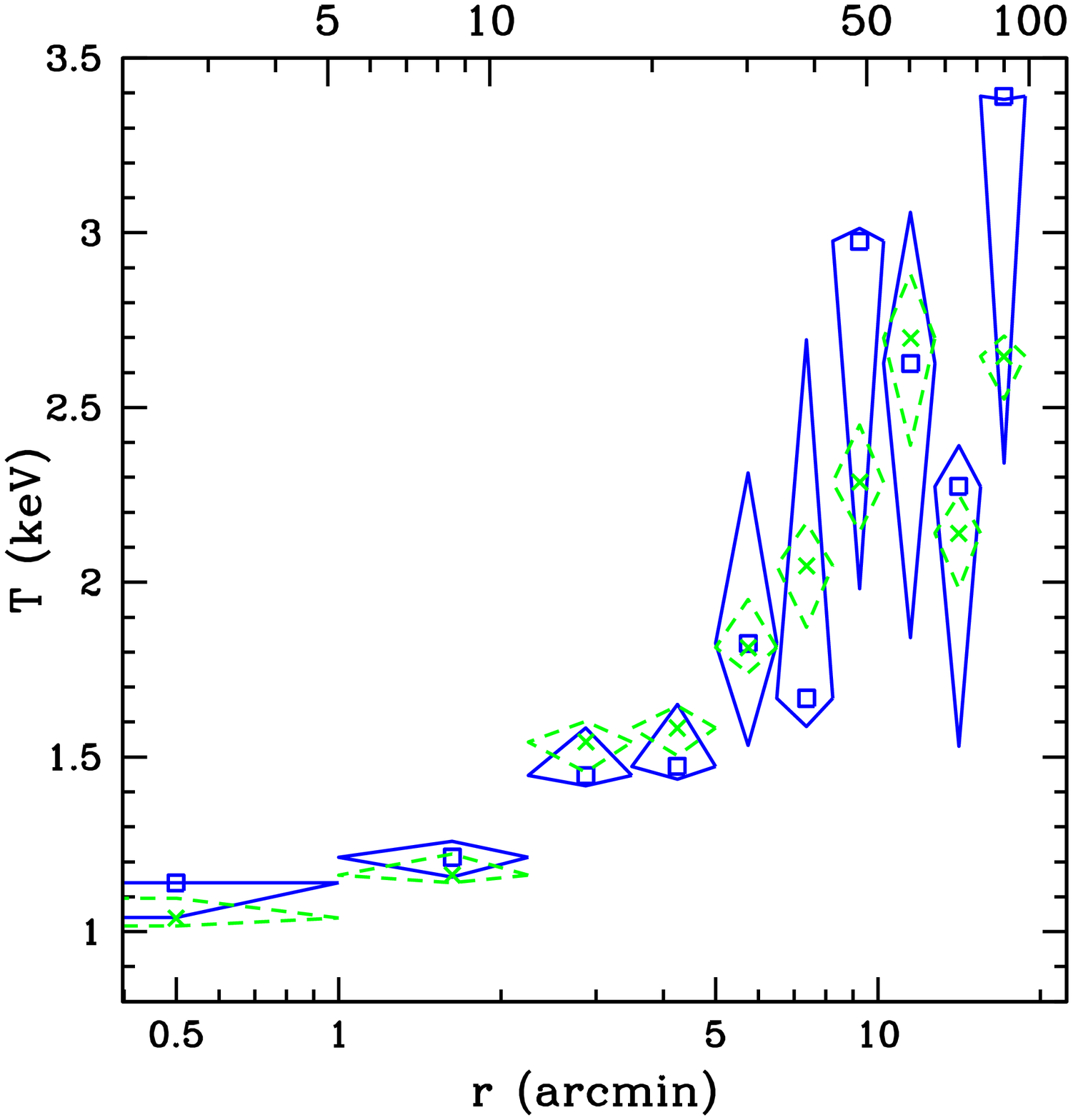,height=0.24\textheight}}
}
\parbox{0.32\textwidth}{
\centerline{\psfig{figure=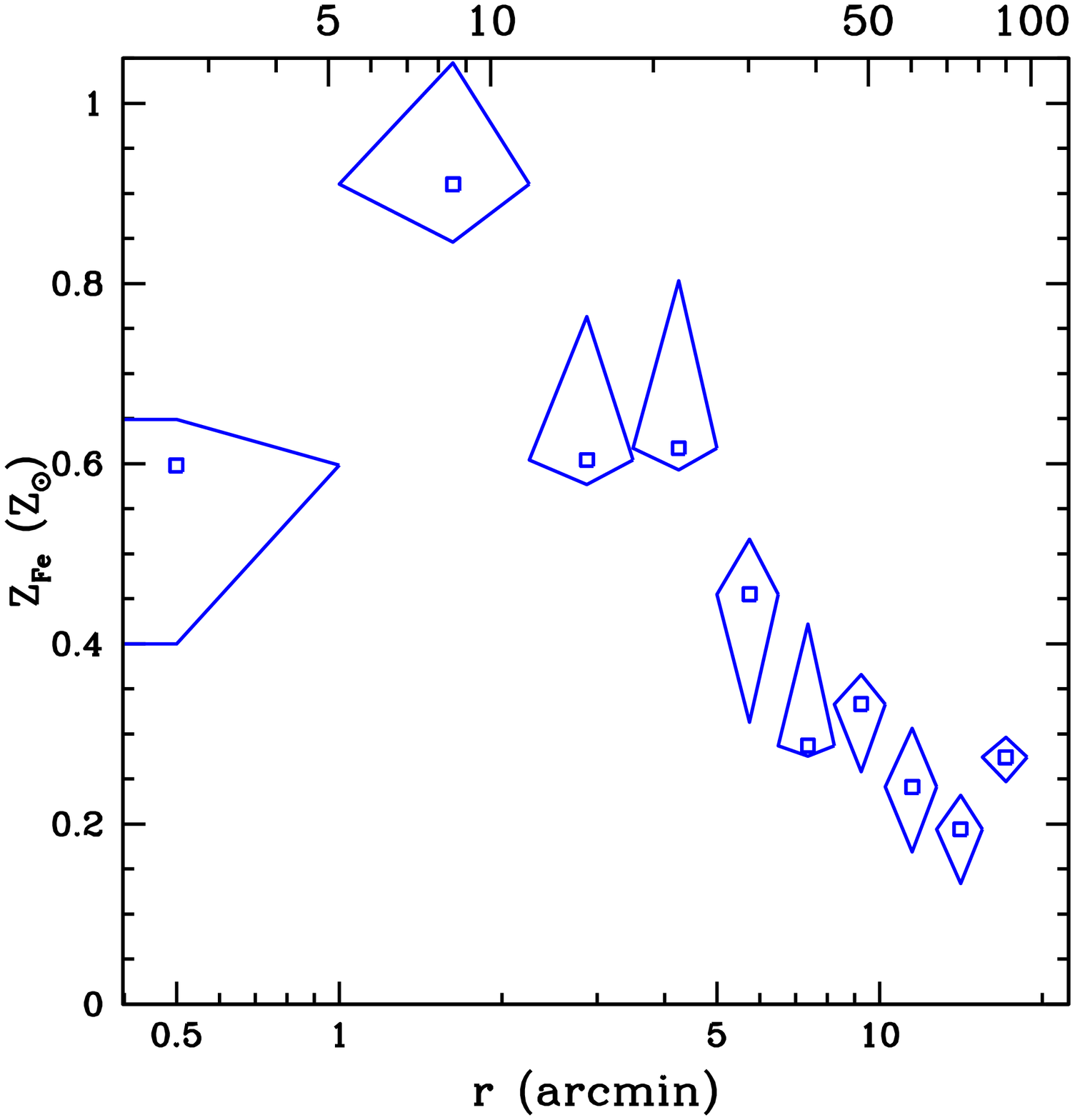,height=0.24\textheight}}
}
\caption{\label{fig.cf} Same as Figure \ref{fig.1t} except for cooling
flow models. Model with fixed \fe and variable \nh\, (1NH\_0Fe\_E) is
denoted by crosses/dashed diamonds/green, and model with Galactic
\nh\, and variable \fe is denoted by boxes/solid diamonds/blue
(0NH\_1Fe\_E). The temperature is associated with the ambient phase
(i.e., $T_{\rm max}$).}
\end{figure*}

\begin{figure*}[t]
\parbox{0.32\textwidth}{
\centerline{\psfig{figure=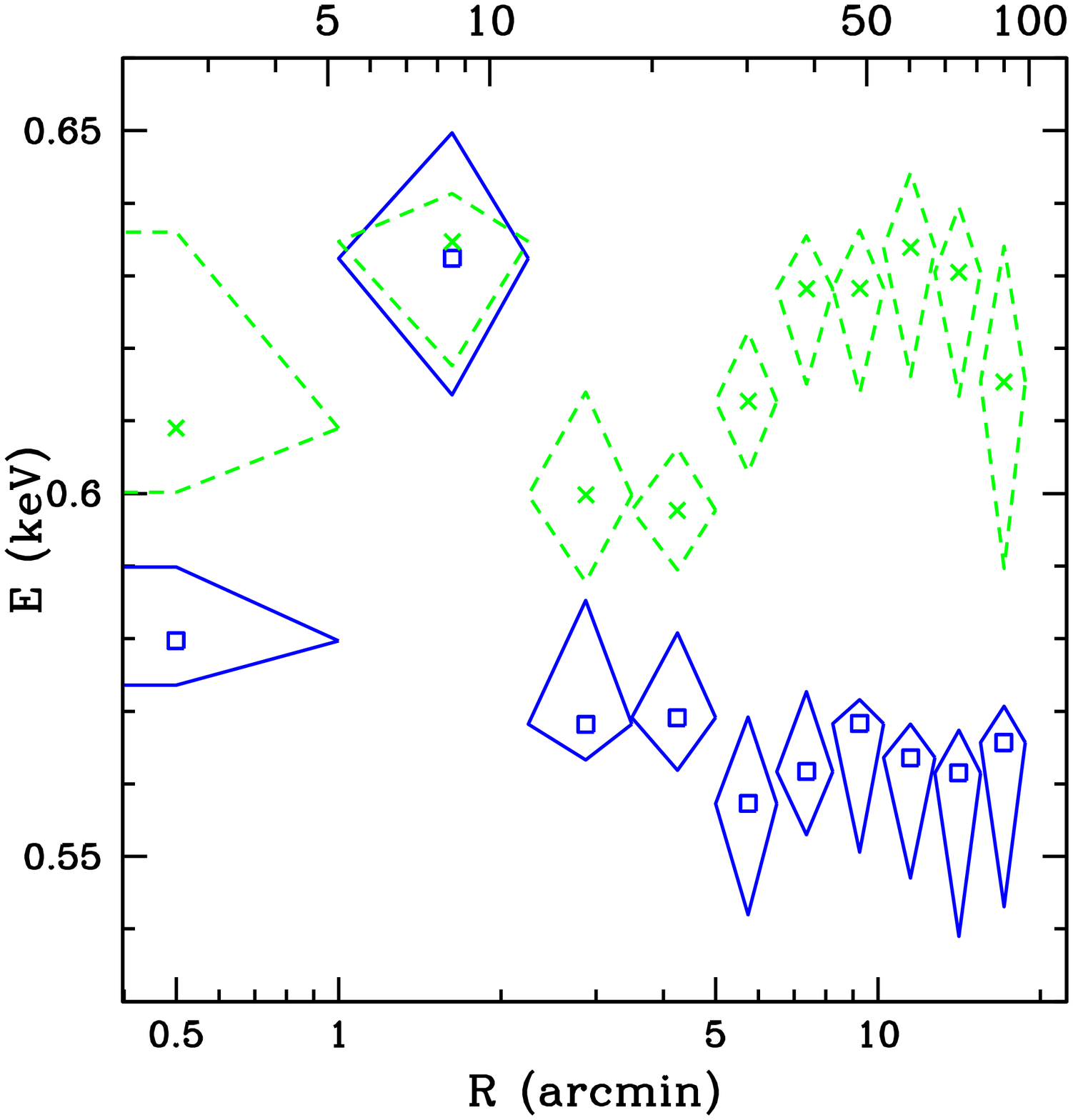,height=0.24\textheight}}
}
\parbox{0.32\textwidth}{
\centerline{\psfig{figure=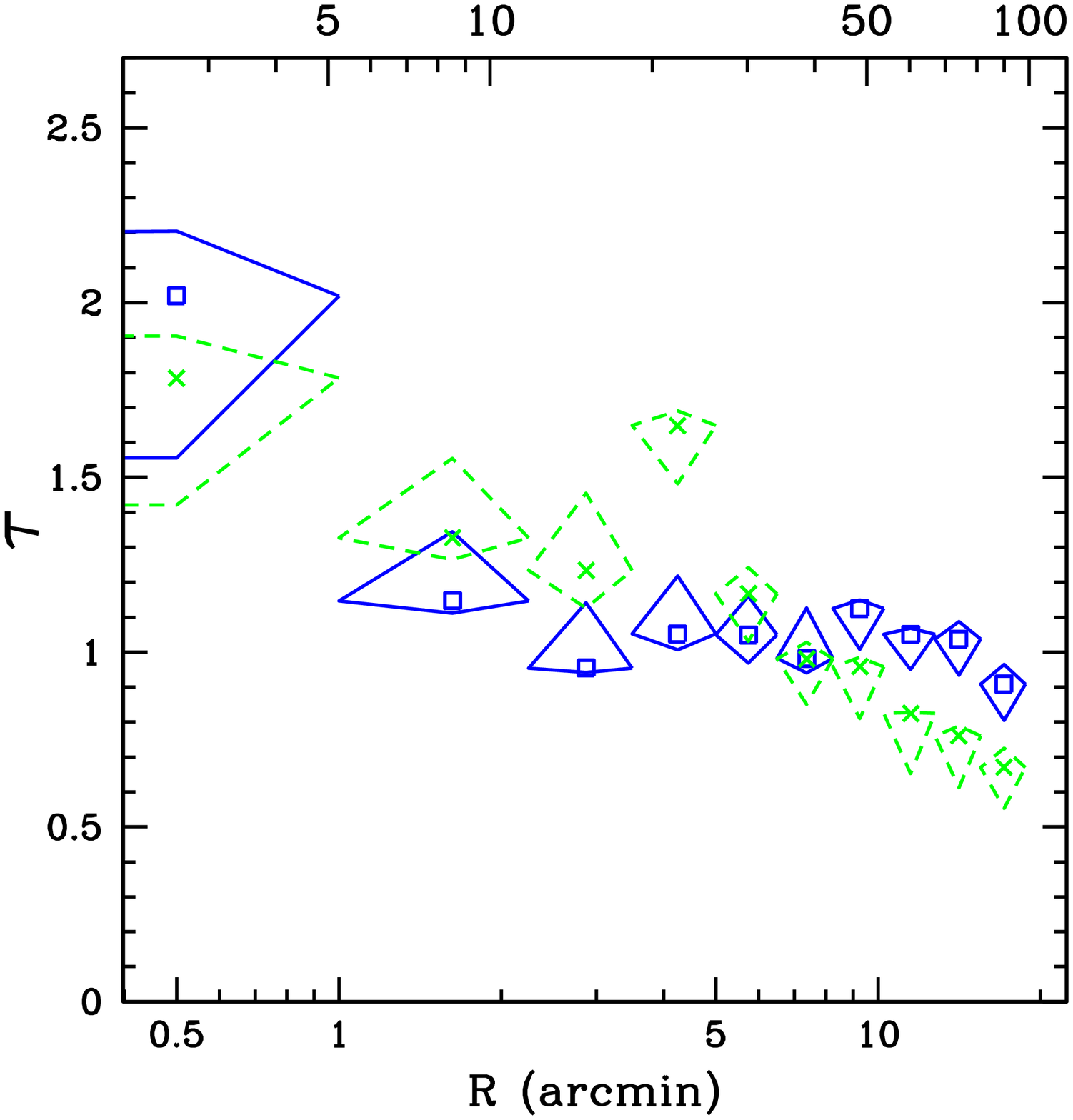,height=0.24\textheight}}
}
\parbox{0.32\textwidth}{
\centerline{\psfig{figure=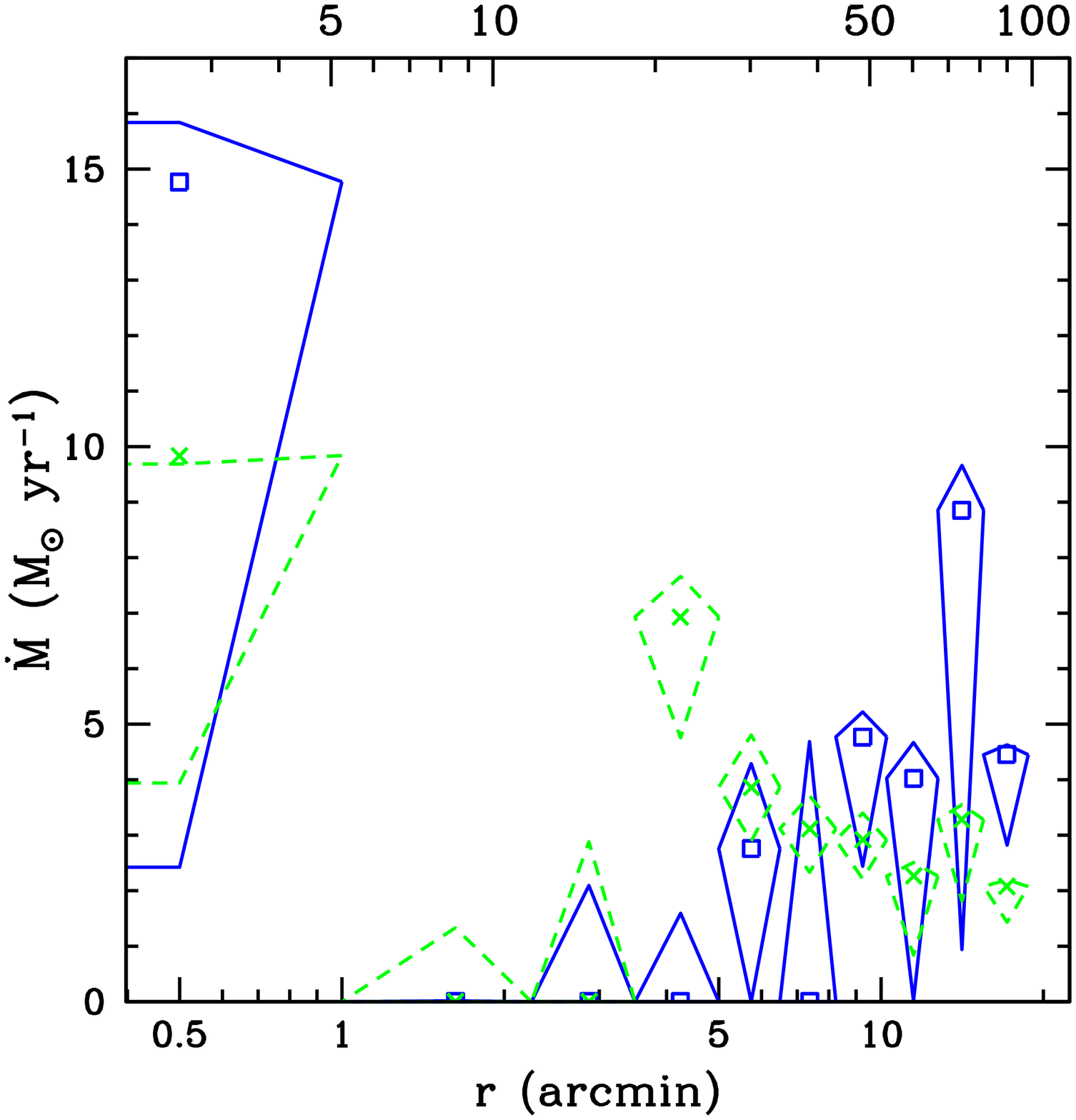,height=0.24\textheight}}
}
\caption{\label{fig.cf_tau} Intrinsic oxygen edge parameters for
cooling flow models: (Left) Oxygen edge energy, $E_0(R)$, for models
1NH\_0Fe\_E (crosses/dashed diamonds/green) and 0NH\_1Fe\_E
(boxes/solid diamonds/blue); (Right) Oxygen edge optical depth,
$\tau(R)$, for same models.}
\end{figure*}

In the inhomogeneous cooling flow scenario the hot gas is expected to
emit over a continuous range of temperatures in regions where the
cooling time is less than the assumed age of the system.  We consider
a simple well-studied model of a cooling flow where the hot gas cools
at constant pressure from some upper temperature, $T_{\rm max}$ (e.g.,
Johnstone et al 1992).  The differential emission measure of the
cooling gas is proportional to $\dot{M}/\Lambda(T)$, where $\dot{M}$
is the mass deposition rate of gas cooling out of the flow, and
$\Lambda(T)$ is the cooling function of the gas (in our case, the
MEKAL plasma code).

Since the gas is assumed to be cooling from some upper temperature
$T_{\rm max}$, the cooling flow model requires that there be a
reservoir of hot gas emitting at temperature $T_{\rm max}$ but is not
cooling out of the flow. Consequently, our cooling flow model actually
consists of two components, CF+1T, where ``CF'' is the emission from
the cooling gas and ``1T'' is emission from the hot ambient gas. We
set $T_{\rm max}$ of the CF component equal to the temperature of the
1T component, and both components are modified by the same
photoelectric absorption. Hence, the cooling flow model adds only one
free parameter $(\mdot)$ to those of the single-phase models.

If only absorption from the standard cold absorber model is considered
then we obtain results identical to the single-phase models in \S
\ref{std}; i.e., the CF component is completely suppressed by the
fits.  Since the CF model includes temperature components below $\sim
0.5$ keV it has stronger \ion{O}{7} ($\sim 0.56$ keV) and \ion{O}{8}
($\sim 0.65$ keV) K$\alpha$ lines than the single-phase models which,
as shown above, already predict too much emission from these
lines. Therefore, in exact agreement with our \rosat analysis of
galaxies and groups in PAPER3 we find that only when an intrinsic
oxygen absorption edge is included in the fits can we obtain a
significant contribution from a CF component, and we shall henceforth
restrict ourselves to cooling flow models with an intrinsic oxygen
edge.

The $\chi^2$ values for selected cooling flow models with an intrinsic
oxygen edge are listed in Table \ref{tab.multi}. Let us focus
initially on the model 0NH\_0Fe\_E for comparison to the single-phase
version. We plot 0NH\_0Fe\_E in Figure \ref{fig.multi_spec} for
annulus \#1 which should be compared to the single-phase version in
Figure \ref{fig.1t_edge_spec}. The cooling flow model in the central
annulus is a much better fit because it can mostly account for the
soft emission over 0.2-0.4 keV. To a lesser extent the cooling flow
model also provides a better fit in annuli \#3-7 and 10, where the
improved $\chi^2$ values are intermediate with those obtained when
varying \nh\, for the single-phase model (1NH\_0Fe\_E in Table
\ref{tab.1t}). Thus, the cooling flow model is able to partially
compensate for the excess soft emission.  (This is also evident from
\nh(R) displayed for model 1NH\_0Fe\_E in Figure \ref{fig.cf}; i.e.,
for $R\la 5\arcmin$ the fitted column densities are consistent with
the Galactic value especially when considering the 5\%-10\%
underestimates arising from the old He cross sections -- see
immediately before \S \ref{std}.)

If \fe is allowed to vary then the quality of the resulting cooling
flow model is very similar to the corresponding single-phase model
(0NH\_1Fe\_E), though the cooling flow gives $\chi^2$ values that are
5-10 lower in most annuli. The temperature and Fe abundance profiles
of the cooling flow model are shown in Figure \ref{fig.cf}. The Fe
abundances are consistent with the single-phase model and the \asca
data \citep{daw00}. However, the temperatures of the cooling flow
model at large radii are 2.5-3.0 keV which, unlike the single-phase
model, are indeed consistent with the \asca/\sax values; i.e., the
cooling flow model 0NH\_1Fe\_E provides slightly better fits and
physical temperatures and Fe abundances in the hot gas in contrast to
the single-phase model.

The absorption parameters obtained for the cooling flow models are
qualitatively similar to those obtained for the single-phase models
(Figure \ref{fig.cf_tau}). The $\tau(R)$ values for 0NH\_1Fe\_E are
larger for the cooling flow model especially at large radii where
$\tau\sim 1.0$ compared to $\tau\sim 0.3$ for the single-phase
version. The larger optical depths for the cooling flow models are
attributed to the need to suppress the stronger \ion{O}{7} and
\ion{O}{8} K$\alpha$ emission lines arising from the cooler
temperature components.

Finally, in Figure \ref{fig.cf_tau} we also plot $\mdot(R)$ for the
two cooling flow models. We see a strange profile where (apparently)
there is significant cooling ($\mdot\sim 10$ \msunyr) at the center
which abruptly terminates in annuli \#2-4 and then rises to a nearly
constant value ($\mdot\sim (3-4)$ \msunyr) at large radius. In the
central bin the value of $\mdot$ is uncertain and consistent with zero
at the 90\% confidence level and therefore is also consistent with the
original \einstein value of $\mdot\sim 3$ \msunyr\, (e.g., Canizares
et al. 1982; Stewart et al. 1984). The total mass deposition rate of
all annuli is $\mdot = 40^{+13}_{-31}$ \msunyr\, for model 0NH\_1Fe\_E
and $\mdot = 34^{+8}_{-14}$ \msunyr\, for model 1NH\_0Fe\_E ($1\sigma$
errors). These values of $\mdot$ are in good agreement with the
previous (imaging) \rosat study by \citet{peres} and are also similar
to the deposition rates of the brightest galaxy groups \citep{b00a}
and small galaxy clusters. The implications of the mass deposition
profile are discussed in \S \ref{feedback}.

\subsection{Two Phases}
\label{2t}

\begin{figure*}[t]
\parbox{0.32\textwidth}{
\centerline{\psfig{figure=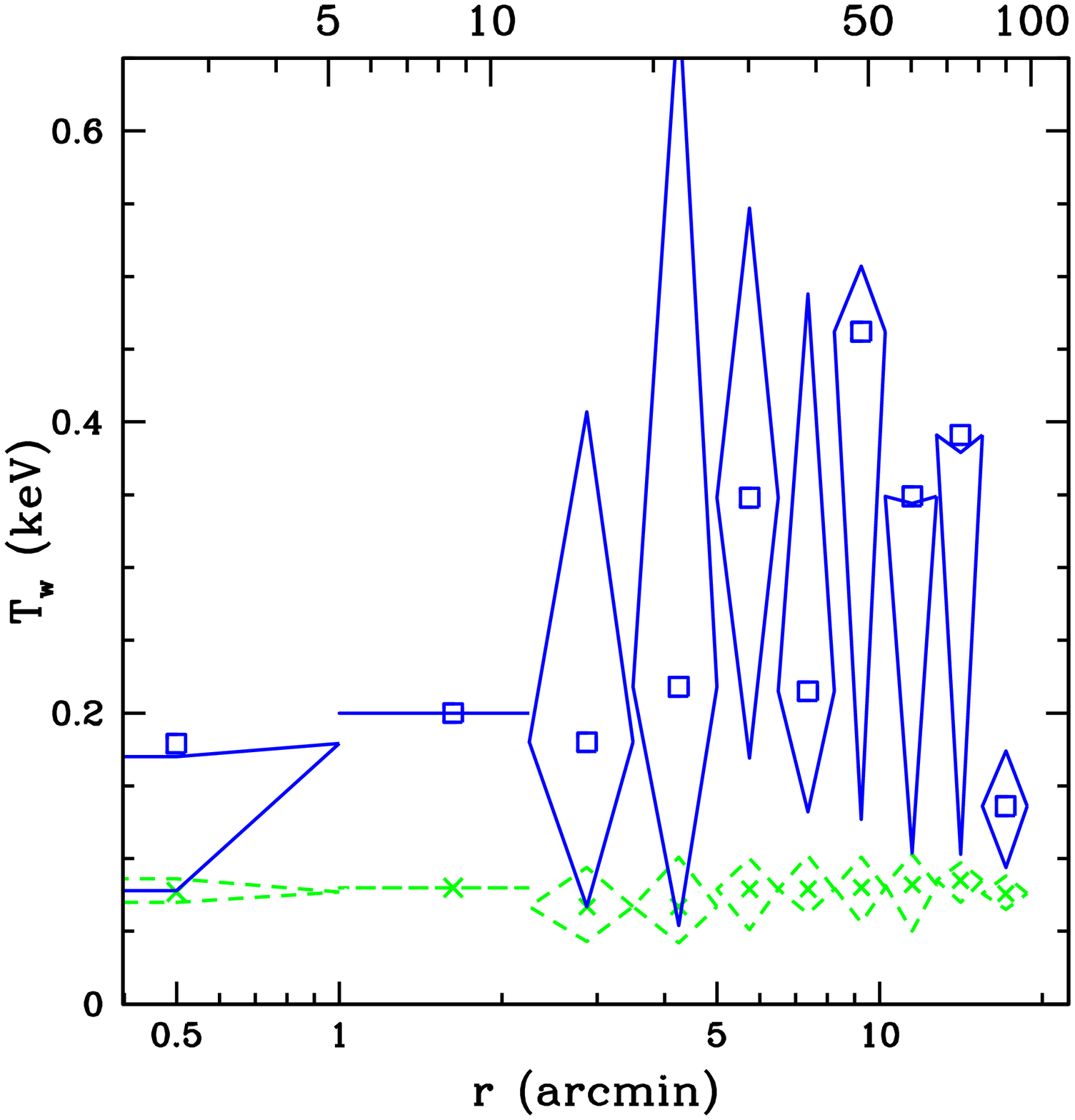,height=0.24\textheight}}
}
\parbox{0.32\textwidth}{
\centerline{\psfig{figure=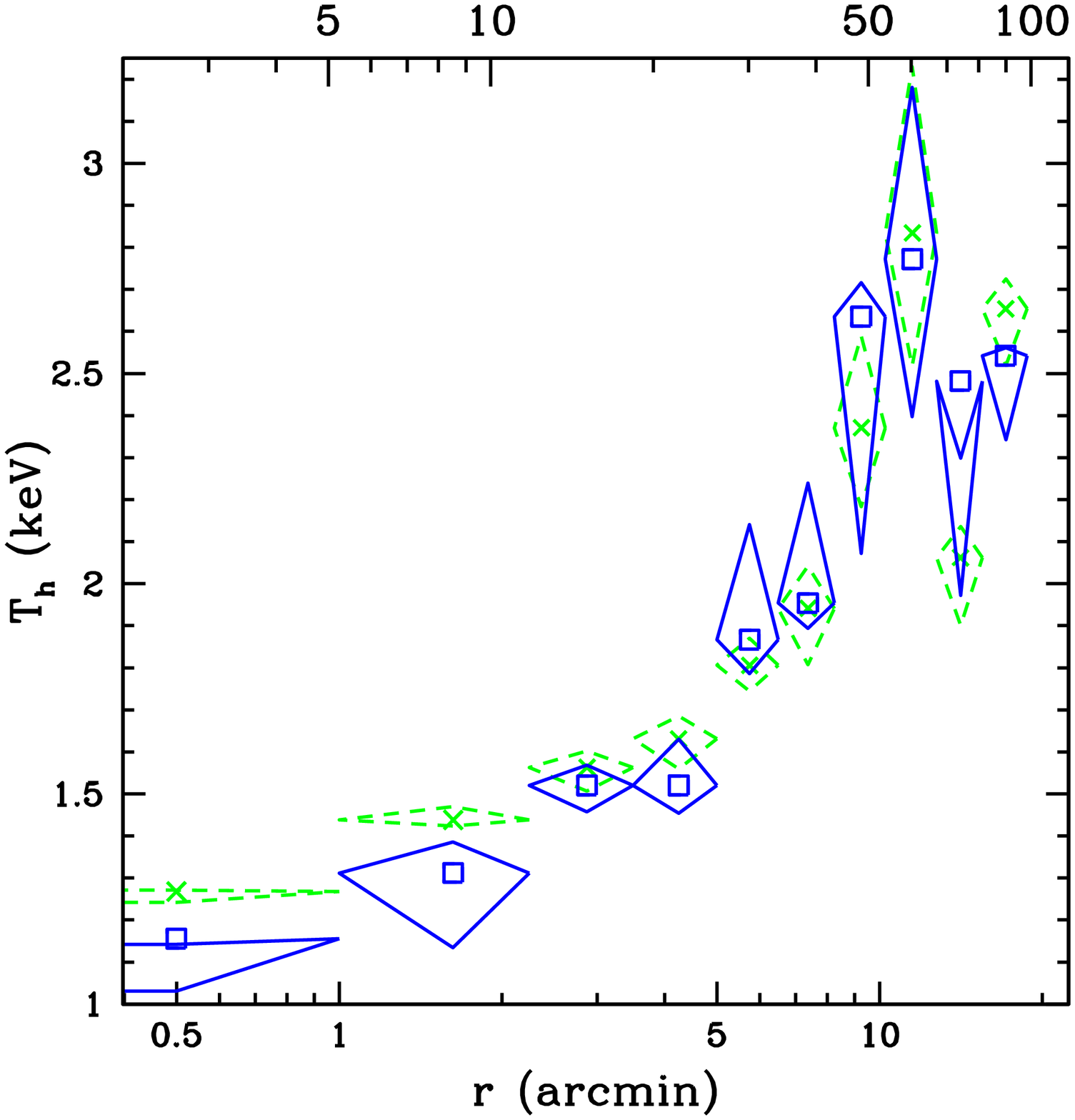,height=0.24\textheight}}
}
\parbox{0.32\textwidth}{
\centerline{\psfig{figure=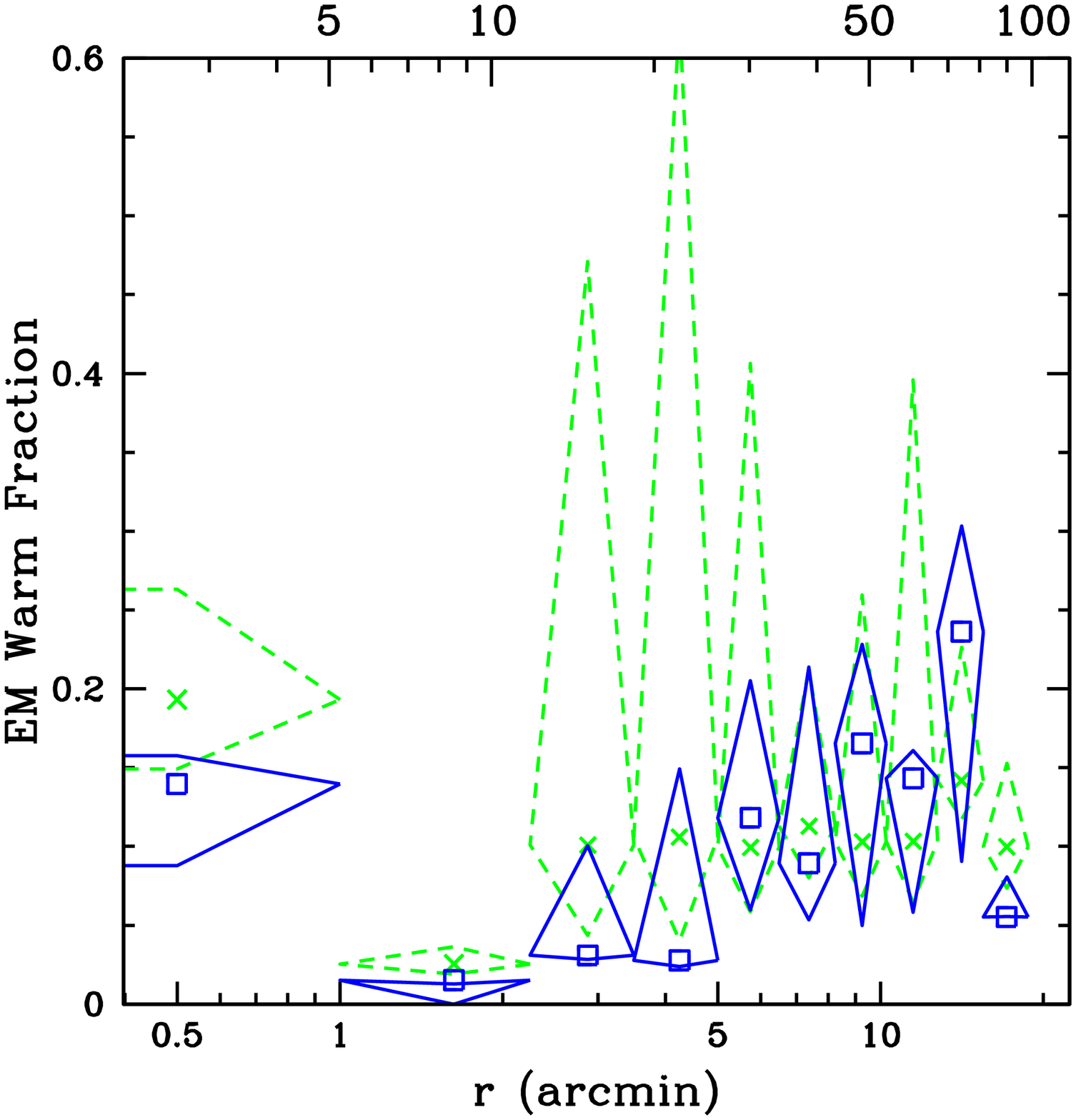,height=0.24\textheight}}
}
\caption{\label{fig.2t} Temperatures for two-phase models without an
intrinsic oxygen edge (0NH\_0Fe in Table \ref{tab.multi};
crosses/dashed diamonds/green) and with the edge (0NH\_0Fe\_E in Table
\ref{tab.multi} ; boxes/solid diamonds/blue). The temperature for the
warm component (\twarm) is shown in the left panel and that of the hot
component (\thot) is shown in the middle panel. The fraction that the
warm component contributes to the total emission measure is shown in
the right panel. Note in each model we fix \twarm\, in annulus \#2 to
the value obtained in annulus \#3 as discussed in the text.}
\end{figure*}

\begin{figure*}[t]
\parbox{0.49\textwidth}{
\centerline{\psfig{figure=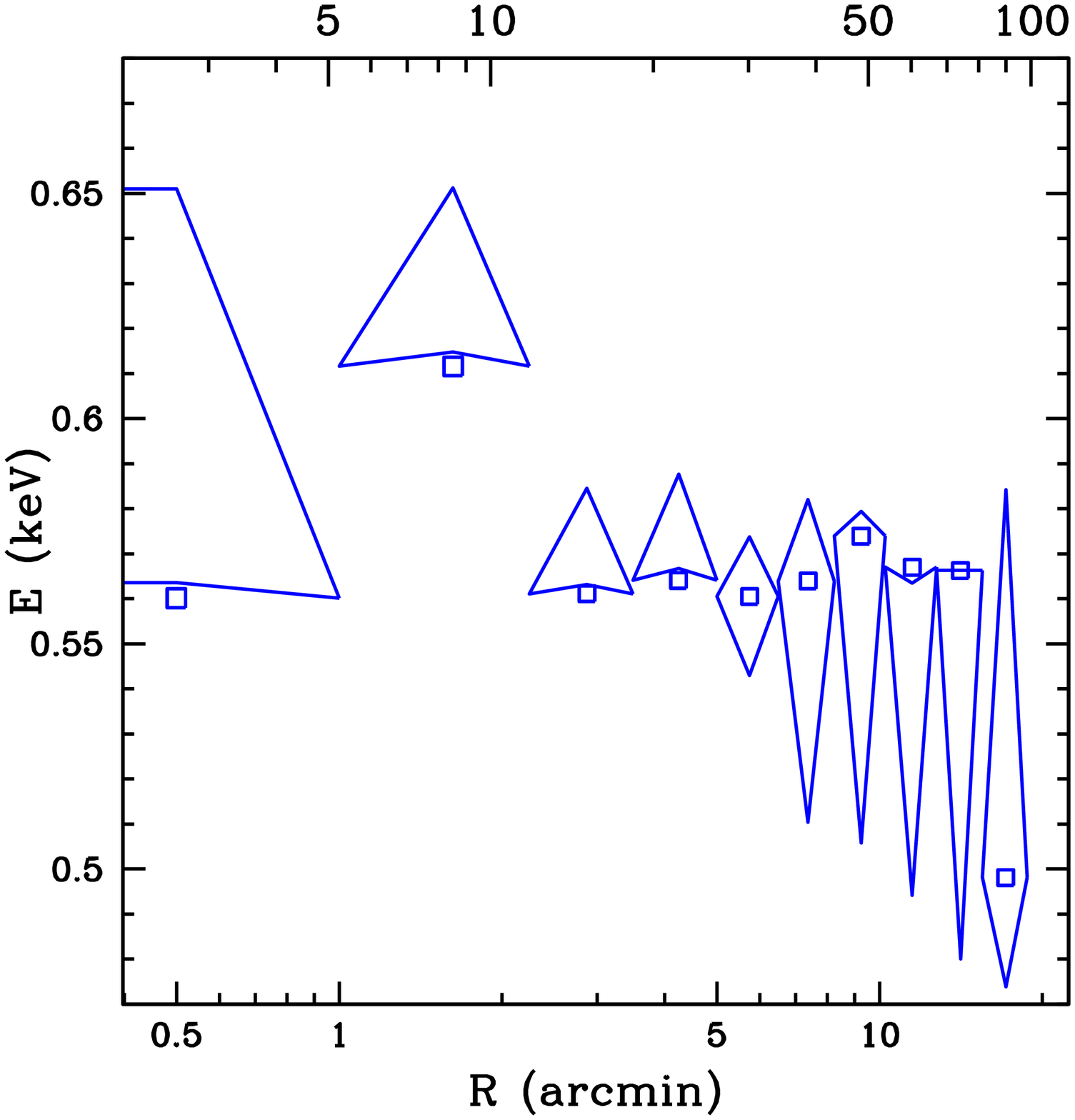,height=0.33\textheight}
}}
\parbox{0.49\textwidth}{
\centerline{\psfig{figure=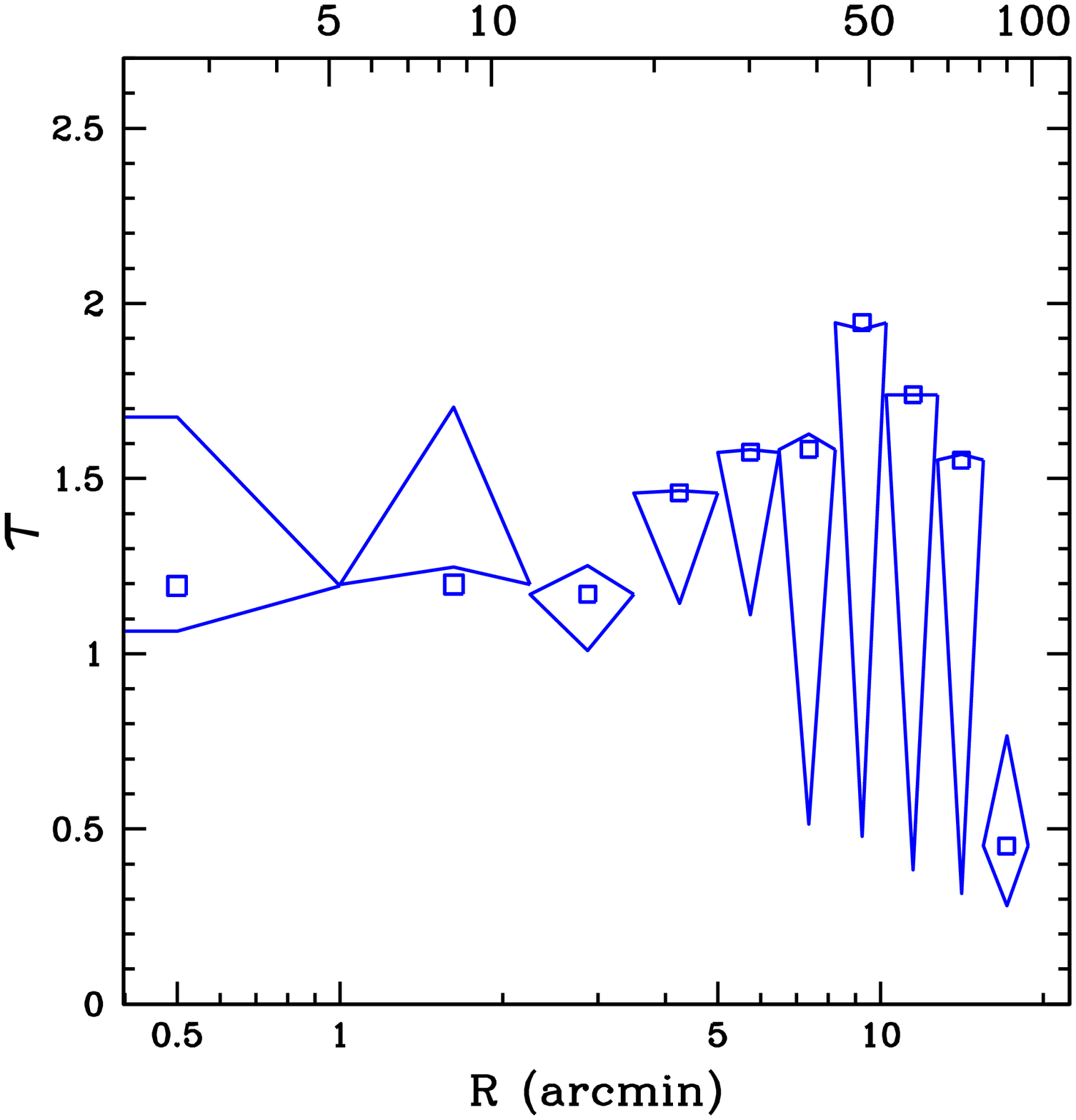,height=0.33\textheight}
}}
\caption{\label{fig.2t_edge_tau} Properties of the intrinsic oxygen
absorption edge for the two-phase model 0NH\_0Fe\_E: (Left) rest
energy and (Right) optical depth.}
\end{figure*}

The constant-pressure cooling flow model is one example of a
(multiphase) gas that emits over a continuous range of temperatures,
and it has the virtue of introducing only one free parameter (\mdot)
to the single-phase models. A two-phase model, however, introduces two
free parameters to the single-phase case: extra temperature and
normalization. From our previous studies (e.g., Buote 1999, 2000a;
Buote et al. 1999) we know that the two-phase model will fit as well
or better than the cooling flow model, and the temperatures and Fe
abundances will be similar in both cases. For our current purposes the
two-phase model is useful since it allows us (1) to assign a specific
temperature to the warm phase and (2) to explicitly quantify the
relative contribution of the warm gas emission to the total.

We expect that adding the ``warm'' temperature component will have a
similar effect as allowing \nh\, to vary in the fits of the
single-phase models. For models without an intrinsic oxygen edge, it
can be seen in Table \ref{tab.multi} that the two-phase models
0NH\_0Fe and 0NH\_1Fe have very similar (actually somewhat better)
$\chi^2$ values than their single-phase counterparts with variable
\nh\, (1NH\_0Fe and 1NH\_1Fe) in Table \ref{tab.1t}.  The same
behavior also applies to the corresponding two-phase models with an
intrinsic oxygen edge (0NH\_0Fe\_E and 0NH\_1Fe\_E). The best-fitting
two-phase model 0NH\_0Fe\_E for annulus \#1 displayed in Figure
\ref{fig.multi_spec} looks quite similar to the corresponding
single-phase version with variable \nh\, (i.e., 1NH\_0Fe\_E) displayed
in Figure \ref{fig.1t_edge_spec}.

The temperature profiles of the two-phase models are shown in Figure
\ref{fig.2t}. The temperature of the warm gas (\twarm) is essentially
constant with radius with value 0.1 keV for the model without the
oxygen edge. For the model with the oxygen edge $\twarm\sim 0.2$ keV
though it can be as large as 0.4 keV at intermediate radii. Although
these temperatures are similar to that of the X-ray background, we
recall that error in the background subtraction cannot account for the
warm emission (\S \ref{std}). Note also that the value of $\twarm$ in
annulus \#2 is very uncertain because the warm emission there is weak,
and thus we had to fix the value of $\twarm$ in annulus \#2 to that of
annulus \#3. The temperature of the hot phase (\thot) in both cases is
similar and is also very consistent with the 2.5-3 keV temperatures
obtained by \asca \citep{daw00} and \sax \citep{dacri} at large radii.

In Figure \ref{fig.2t} we also show the fraction that the warm
temperature component contributes to the total emission measure. Over
most radii the warm temperature component contributes 10\%-20\% to the
emission measure. However, in annulus \#2 the warm component is
markedly suppressed and is consistent with zero for the model with an
oxygen edge. The variation of emission-measure fraction with radius is
qualitatively similar to the profile of \mdot\, for the cooling flow
model (Figure \ref{fig.cf}); i.e., even though the cooling flow and
two-phase models have entirely different differential emission
measures they give essentially identical pictures of the variation of
warm gas with radius.

The absorption parameters of the intrinsic oxygen edge for the
two-phase model 0NH\_0Fe\_E are displayed in Figure
\ref{fig.2t_edge_tau}. Both the rest energies of the edge and the
optical depths are (nearly) consistent with being constant with
radius. Although these flat profiles are somewhat different from the
cooing flow and single-phase models, the ranges of values of energies
and optical depths are similar. 

\section{Discussion}
\label{disc}

\subsection{Comparison with Previous X-Ray Studies of M87}
\label{previous}

Using a spatial-spectral deprojection analysis we have shown that a
single-phase model of the hot gas modified by standard Galactic
absorption cannot describe the \rosat PSPC data of M87. The failure of
the single-phase model is dramatically evident in the spectral plot in
Figure \ref{fig.1t_spec}: excess absorption over 0.5-0.8 keV and
excess emission over 0.2-0.4 keV signal the presence of an additional
warm ($\sim 10^6$ K) gas component.  This evidence for warm absorbing
(and emitting) gas in M87 lends strong support to the results we have
recently obtained from \rosat data of several bright galaxies and
groups (PAPER1 and PAPER3).

Only multiphase models (e.g., cooling flow, two-phase) with an
intrinsic oxygen edge are able to provide good fits to the \rosat data
and still have (1) ambient temperatures and Fe abundances consistent
with \asca and \sax studies (\S \ref{std}) and (2) Galactic hydrogen
column densities for the known absorption arising from the intervening
cold gas in the Milky Way. Although the multiphase models provide
better fits than the single-phase models, they still do not provide
perfect fits in all annuli (Figure \ref{fig.multi_spec}). The
remaining low-level residuals can be eliminated by allowing the
$\alpha$-process elements to vary differently from Fe; e.g., the most
important element is Si which removes most of the residual if Si/Fe is
allowed to be several times solar. A modest super-solar Si/Fe ratio is
consistent with previous \asca studies (Buote et al. 1999 and
references therein). Since \rosat is unable to place tight constraints
on the abundance ratios we do not give constraints on these quantities
at this time.

It is also possible that these low-level residuals in our best models
reflect calibration errors. In particular, the remaining feature near
0.3 keV (Figures \ref{fig.1t_edge_spec_nh} and
\ref{fig.multi_spec}) is very similar to that seen in the PSPC spectra
of some other bright clusters in the recent study by
\citet{ab00}. These authors prefer to explain these low-level
residuals as a gain offset.

While preparing our manuscript we became aware of the preprint by
\citet{sfa}. These authors also present a spatial-spectral
deprojection analysis of the \rosat data of M87 along with a large
sample of clusters. The analysis of \citet{sfa} differs from ours in
two main respects. First, they bin up each PSPC spectrum into three
X-ray channels (or ``colors''). Spectral models are then fitted to
ratios of these channels. Second, they exclude data below 0.4 keV and
ignore the sensitivity of the inferred \nh\, to
\emin\, shown in our Figure \ref{fig.nhvsemin}. Despite these
differences between our respective analyses, \citet{sfa} also conclude
that a single-phase model is inadequate and that spatially extended
intrinsic absorbing material is required. Since, however, \citet{sfa}
exclude data below 0.4 keV they do not detect the excess soft emission
and thus are not led to conclude that the absorber must be warm.

Intrinsic absorption in M87 is also inferred from two-temperature
models of the integrated \asca spectra within $R\sim 5\arcmin$ by
\citet{adf} which confirms the original detection with \einstein
\citep{daw91}.  The hydrogen column density of cold material implied
by the two-temperature \asca models, $(4-5)\times 10^{21}$ cm$^{-2}$,
is considerably larger than the value of $1.4\times 10^{21}$ cm$^{-2}$
obtained by \citet{sfa} with \rosat data. The smaller value obtained
by \citet{sfa} arises from their fitting \rosat data with energies
above 0.4 keV with a cold absorber whereas the \asca analysis of
\citet{adf} includes only data above 0.6 keV.  As we have shown in
Figure \ref{fig.nhvsemin} the inferred value of \nh\, is highly
sensitive to the lower energy limit of the bandpass.  Our absorption
models using a single oxygen edge yield optical depths of $\sim 1$ in
the inner regions of M87. These optical depths translate to
$\nh=(3-4)\times 10^{21}$ cm$^{-2}$ (depending on the ionization state
of oxygen) which agree with the \asca result.

The new \rosat evidence for multiphase gas in M87 largely confirms and
extends the original detection by \citet{crc82} within the central
$\sim 2\arcmin$ using the \einstein FPCS. By measuring selected line
blends of \ion{Fe}{17} - \ion{Fe}{24} \citet{crc82} inferred the
presence of plasma emission components with temperatures ranging from
$3\times 10^7$ K down to $2-4\times 10^6$ K consistent with our
results. However, these authors also deduce from analysis of the
ly-$\alpha$ \ion{O}{8} line that the O/Fe ratio is 3-5 times solar in
the ISM (standard Galactic absorption was also assumed). Since this is
not confirmed by either the \rosat or \asca data (which instead
indicate that the data are absorbed over 0.5-0.8 keV where the
ly-$\alpha$ \ion{O}{8} line emits) the super-solar O/Fe ratio probably
can be attributed to the FPCS calibration errors discussed by
\citet{tsai} for M87.

An alternative explanation (due to the referee) for the FPCS
measurement is as follows. Since the FPCS operated over a narrow
energy band, if the absorption edges are confined to energies just
below the \ion{O}{8} K$\alpha$ line they would not be seen by the
FPCS. Consequently, the FPCS would measure the line strength with
respect to the absorbed continuum level. Lower resolution detectors
like \rosat and \asca would merge the edges and line and therefore see
a smaller line strength.

\subsection{Mass of the Warm Gas}
\label{mass}

We now consider how much mass is associated with the warm ionized
absorber in M87. In principle this can be computed from the optical
depths of the intrinsic oxygen edges (Figures \ref{fig.1t_edge_tau},
\ref{fig.cf_tau}, and \ref{fig.2t_edge_tau}), the mass deposition
rates of the cooling flow models (Figure \ref{fig.cf_tau}), or the
emission measures of the warm gas in the two-phase models (Figure
\ref{fig.2t}). Unfortunately, our single-edge absorption model is
really just a phenomenological tool to establish the existence and to
study the gross properties of the absorber which is appropriate for
the low resolution of the PSPC (\S \ref{edge}). As discussed in
PAPER3, when including additional edges the inferred optical depth for
each edge decreases, and we expect several edges from different
ionization states of oxygen (also carbon and nitrogen) to
contribute. Since the absorption of an edge is not linear in the
energy of the edge (i.e., $A_0(E)=\exp[-\tau_0(E/E_0)^{-3}]$ and
$A_1(E)A_2(E) \ne A_{1+2}(E)$ if the edge energies $E_1 \ne E_2$), by
spreading multiple edges over a large energy range one can produce the
observed absorption with smaller total optical depth than can be
achieved with a single edge.  Consequently, the single-edge optical
depths should be considered upper limits.

Nevertheless, to provide a convenient benchmark for comparison to
PAPER3 and to related previous studies, we have computed the absorber
masses by assuming the absorption edges correspond to \ion{O}{1} (see
\S 5.3 of PAPER3). Using the single-phase model 1NH\_0Fe\_E as an
example, we find that within the central bin (5 kpc) that the absorber
mass is, $\mabs= 3.3_{- 0.8}^{+ 0.3}\times 10^9\msun$ ($1\sigma$
errors) which is about twice that of the hot gas, $\mhot= 1.20_{-
0.05}^{+ 0.01}\times 10^9\msun$. The total values within 98 kpc are
respectively, $\mabs= 2.4_{- 0.3}^{+ 0.3}\times 10^{11}\msun$ and
$\mhot= 2.14_{- 0.04}^{+ 0.02}\times 10^{11}\msun$; i.e.,
$\mabs\approx \mhot$ similar to that found for the galaxies and groups
in PAPER3.  The amount of time required to accumulate $\mabs$ assuming
$\mdot = 40^{+13}_{-31}$ \msunyr (\S \ref{cf}) is $\tacc = 0.6_{-
0.2}^{+ 2.4}\times 10^{10}\rm yr^{-1}$; i.e., this absorbing material
could reasonably have been accumulated by a cooling flow in agreement
with the galaxies and groups in PAPER3.

If, as we expect, the absorber is the same material that gives rise to
the excess 0.2-0.4 keV emission, then the above values of \mabs\,
should be (nearly) the same as $M_{\rm warm}$; i.e., the mass of the
warm gas component of the two-phase models (\S \ref{2t}).  However,
there are two key problems with translating the results of \S \ref{2t}
into values of $M_{\rm warm}$. First, the temperatures of the warm
components are 0.1-0.2 keV, and it has recently been emphasized that
the current plasma codes (including MEKAL) can underestimate the
emission around 0.2 keV by as much as a factor of 10 at these low
temperatures because of neglected line emission (Liedahl 1999;
Beiersdorfer et al. 1999). Second, since the warm gas absorbs photons
from the hot gas the assumption of collisional ionization equilibrium
is also not strictly valid nor is it clear that the warm gas is fully
optically thin to its own radiation as we have assumed for convenience
(with MEKAL).

These problems are evident when we attempt to compare the emission
measures ($\propto \rm n_e^2V)$ of the warm gas component in the
two-phase model to that implied by the values of \mabs\, quoted above
from the oxygen edges. We find that, e.g., in the central radial bin
the emission measure implied by the oxygen edges is $\sim 30$ times
that inferred for the warm gas component in the two-phase model. This
discrepancy is probably exacerbated by the ``missing line'' problem in
the plasma codes, but can be mitigated by consideration of the
following. First, by including multiple edges in our fits we can
reduce the total optical depth by factors of a few leaving a factor of
$\sim 10$ discrepancy. Second, if we consider the radiative transfer
effects mentioned above (i.e., photoionization) then the warm gas is
over-ionized for its (collisional) temperature. The inferred
(collisional) emission measure is very sensitive to the temperature,
and if we reduce the 0.1-0.2 keV temperature by a factor of $\sim 3$
the discrepancy of emission measures can be fully mitigated.

This exercise highlights the need for a more rigorous (and complex)
treatment of the absorption and emission of the warm gas in order to
deduce the mass of the absorber which is beyond the scope of our
paper. If our basic results are confirmed with the substantially
higher quality data from \chandra and \xmm, then it will be
appropriate to expend the effort to construct rigorous models of the
warm absorber accounting for many edges and possible radiative
transfer effects that are not currently available in \xspec.

Finally, it is also worth mentioning that (these caveats aside) the
emission from the warm gas implied by our measurements does not
violate published constraints in the FUV similar to that found for the
the galaxies and groups in section 5.4 of PAPER3. \citet{dhf} have
reported results in the FUV for M87 obtained with the Hopkins
Ultraviolet Telescope (HUT). They obtain a $2\sigma$ upper limit of
$\sim 1\times 10^{-10}$ \ergcms within $R=1\arcmin$ of M87 for the
flux of the \ion{O}{6} (1034\AA) line which is expected to be the
strongest line associated with warm gas having temperatures between
$10^{5-6}$ K. Using \mabs\, quoted above we estimate that the strength
of the \ion{O}{6} line at the peak temperature of $3.2\times 10^5$ K
(e.g., Pistinner \& Sarazin 1994; Voit \& Donahue 1995) is $\sim
9\times 10^{-10}$ \ergcms. Ignoring the fact that \mabs\, is
over-estimated as discussed above, the HUT constraint implies that the
\twarm\, must be different from $3.2\times 10^5$ K. From examination
of Figure 7 of \citet{vd} we see that the \ion{O}{6} luminosity
decreases very rapidly with increasing temperature. Consequently, the
emission implied by \mabs\, is consistent with the HUT constraint
provided that $\twarm\ga 5\times 10^5$ K.

\subsection{EUV Emission}
\label{euv}

Various conflicting reports of the detection of excess EUV (0.07-0.25
keV) emission above that expected from the hot (X-ray) plasma have
been reported in clusters of galaxies using data from the {\sl Extreme
Ultraviolet Explorer} (\euve), but only for M87 does there appear to
be a consensus among different groups of the reality of centrally
concentrated excess EUV emission (Lieu et al. 1996; Bergh\"{o}fer et
al. 2000).  If the excess \euve emission is interpreted as arising
from optically thin collisionally ionized gas then \citet{lieu96} and
\citet{berghofer} infer temperatures ($\sim 0.1$ keV) and total masses
within $R=19\arcmin$ ($\sim 1\times 10^{11}\msun$), which (considering
the modeling differences between our studies) are in good agreement
with those values we have inferred from the \rosat data.

We mention that it has usually been reported that the \euve ``soft
excess'' of M87 increases with radius (e.g, Bonamente, Lieu, \& Mittaz
2000). This result depends on the definition of the model of the hot
gas that serves as the reference for the excess. The radially
increasing soft excess assumes that the reference model for the hot
gas has been fitted to the PSPC data using only data above 1 keV. If
instead all the PSPC data are included in the fit, then there is no
significant trend for the soft excess with radius in the \euve data
(R. Lieu 2000, private communication). This is consistent with our
result for the two-phase models where the warm gas fraction is
essentially constant for $R\ga 5\arcmin$ (Figure \ref{fig.2t}).

Various problems with a gas origin of the EUV emission have been
discussed by \citet{acf96}, in particular the need for some
as-yet-undiscovered heat source to prevent the gas from rapidly
cooling to lower temperatures. These problems have spurred many to
explain the excess EUV emission from other non-thermal processes (see
Bergh\"{o}fer et al. 2000 and references therein). However, an
independent test of the hypothesis of warm ionized gas is offered by
its absorbing effect on the hot X-ray plasma. Our detection of
absorption confined to energies 0.5-0.8 keV is precisely the signature
expected from warm ($\sim 0.1$ keV) gas (e.g., Krolik \& Kallman
1984).

The results from \euve, \rosat, and \asca for M87 form a coherent
picture of a multiphase medium consisting (approximately) of a warm
and hot phase: The emission from the warm gas is detected with \euve
and \rosat while the absorption from the warm gas is detected in
\rosat and \asca.

\subsection{Implications for Cooling Flows and AGN Feedback}
\label{feedback}

The warm ($\sim 10^6$ K) gas we have detected in both absorption and
emission with \rosat, which is consistent with both the excess
absorption seen in \asca and the excess emission in \euve, provides
direct evidence for the material that may have cooled and dropped out
of an inhomogeneous cooling flow (e.g., Fabian 1994). Although the
warm gas is consistent with the amount of matter that could have been
deposited by a cooling flow (\S \ref{mass}), it is difficult to
explain the warm gas phase as an equilibrium configuration because (1)
the gas at $T\sim 10^6$ K cannot be thermally supported in the
potential of M87, and (2) the cooling time of this gas is very short
(see section 5.5 of PAPER3).

Non-equilibrium models in the form of feedback from the central AGN
have been proposed to largely suppress cooling flows (Binney \& Tabor
1995; Ciotti \& Ostriker 2000). In the feedback model whenever the
black hole accretes a sufficient amount of gas to stimulate nuclear
activity, the accompanying radiation stimulated by the accretion heats
up the hot gas and prevents further cooling. This is supposed to be a
cyclical process such that the AGN phase is sufficiently rare to be
consistent with the lack of nuclear activity in most cooling flows.

The spectacular radio map of M87 published by \citet{owen} shows that
radio emission pervades the hot gas of M87 out to radii of $\sim
5\arcmin$. The distortions in the \rosat X-ray contours present in
annuli \#2-3 (Figure \ref{fig.m87}) testify to the interaction of the
radio emission associated with the jet and the hot gas \citep{harris}.
The interpretation of the radio emission of M87 in terms of the AGN
feedback model has been discussed by \citet{binney99}.

Our multiphase models of the \rosat PSPC data of M87 appear to support
the picture of a standard inhomogeneous cooling flow that has been
partially disturbed by the AGN. We focus the reader's attention on the
radial profiles of both \mdot\, for the cooling flow model in Figure
\ref{fig.cf} and the emission-measure fraction of warm gas in
two-phase models shown in Figure \ref{fig.2t}. In particular, for the
cooling flow models \mdot\, is consistent with a constant value for
$R>5\arcmin$ appropriate for a standard multiphase cooling flow. But
for annuli \#2-4 (i.e., $R=1\arcmin$ to $R\sim 4\arcmin$) \mdot\, is
zero. Over this region lies the asymmetrical features in the X-ray
isophotes and the most prominent radio emission from the jet which
would seem to support the notion that the AGN has disturbed the
cooling flow in this region.  In the central bin ($R\le 1\arcmin$) we
do see significant cooling (especially for the two-phase models) which
is not surprising since the relaxation and cooling times are shortest
there, and thus it would be the first to readjust after the
disturbance from the AGN.

It should be remembered that the scenario we have described has been
based on the {\it emission} component of the warm gas. The absorption
given by the optical depth of the single edge is essentially constant
(or declining slowly) with increasing radius. Consequently, although
the AGN may have inhibited the cooling of the warm gas in the central
regions it apparently has not affected much the absorption properties
of the gas. Whether or not this is reasonable requires a more complex
model of the warm absorber/emitter than we have employed in this paper
(see \S \ref{mass}).  If this description is correct for M87 then it
very likely applies to other cooling flows. In this case it will be
critical to understand how this process evolves and how ubiquitous it
is in cooling flow galaxies, groups, and clusters.

\section{Conclusions}
\label{conc}

From analysis of spatially resolved, deprojected \rosat PSPC spectra
we find the strongest evidence to date for intrinsic oxygen absorption
and multiphase gas in the hot ISM/IGM/ICM of a galaxy, group, or
cluster. When attempting to describe the 0.2-2.2 keV \rosat emission
of M87 by a single-phase hot plasma modified by standard Galactic
absorption the best-fitting model displays striking residuals in the
spectrum (Figure \ref{fig.1t_spec}): (1) excess emission above the
model for 0.2-0.4 keV and (2) excess absorption below the model for
0.5-0.8 keV. These features are apparent out to the largest radii
investigated ($\sim 100$ kpc) and cannot be attributed to errors in
the calibration or the background subtraction (see \S \ref{std}).

The principal result is that the 0.5-0.8 keV absorption is consistent
with that of a collisionally ionized plasma with a temperature of
$10^{5-6}$ K, where the lack of evidence of absorption below 0.5 keV
strongly excludes the possibility of absorption from cold material as
has been assumed in essentially all previous studies of absorbing
material in cooling flows. In fact, the excess {\it emission} observed
between 0.2-0.4 keV, which is also manifested as sub-Galactic column
densities in models with standard (cold) absorbers (\S \ref{single}),
has a temperature that is consistent with the emission from gas
responsible for the 0.5-0.8 keV absorption and could not be explained
if dust were responsible for both the absorption and soft
emission. (See section 5.1 of PAPER3 for discussion of other problems
with dust.)

Only multiphase models can provide good fits over the entire PSPC
bandpass while also yielding temperatures and Fe abundances that are
consistent with results from \asca and \sax at large radii (\S
\ref{multi}). Both cooling flow and two-phase models indicate that the
fraction of warm gas with respect to the total emission measure
differs qualitatively for radii interior and exterior to $\sim
5\arcmin$ ($\sim 26$ kpc). For $r>5\arcmin$ the data are consistent
with a constant (or slowly varying) fraction of warm gas as a function
of radius. But for $r<5\arcmin$ the warm gas fraction varies from
$\sim 20\%$ within the central bin $(1\arcmin)$ to essentially zero
within the next few bins (out to $5\arcmin$). This behavior is to be
contrasted with the absorption optical depth profiles which are
approximately constant with radius.

The oxygen absorption and soft emission from warm ($10^{5-6}$ K) gas
in M87 we have detected using the \rosat PSPC (0.2-2.2 keV) is also
able to satisfy the detections of excess emission with data at lower
energies (0.07-0.25 keV) from \euve (\S \ref{euv}) and the detection
of excess absorption at higher energies (0.5-10 keV) from \einstein
and \asca (\S \ref{mass}). Previous studies using only \euve data
(e.g., Bergh\"{o}fer et al. 2000) could not decide between a
non-thermal and thermal origin for the excess soft emission.
Similarly, the previous detections of absorption with \einstein
\citep{daw91} and \asca \citep{adf} could not constrain the
temperature of the absorber and always assumed a cold absorber with
solar abundances. Even previous studies with \rosat that neglected
data below $\sim 0.5$ keV also could not constrain the temperature of
the absorber (Allen \& Fabian 1997; Sanders et al. 2000) because of
the strong dependence of the inferred absorption on the lower energy
limit of the bandpass (see our Figure \ref{fig.1t_spec}).

Hence, the \rosat detection of intrinsic absorption that is localized
in energy (0.5-0.8 keV) is the key piece of evidence for establishing
the presence of warm ($10^{5-6}$ K) gas distributed throughout (at
least) the central 100 kpc. This evidence for a multiphase ISM in M87
essentially confirms the original detection within the central $\sim
2\arcmin$ using the \einstein FPCS \citep{crc82}. However, instead of
intrinsic oxygen absorption \citet{crc82} inferred a super-solar O/Fe
ratio which does not agree with subsequent analyses of M87 using other
instruments. The anomalous O/Fe ratio is probably attributed to
either calibration error in the FPCS (see Tsai 1994) or to an
underestimate of the continuum due to the absorption edges (see end of
\S \ref{previous}).

Although many previous studies have shown that the \rosat data
(0.1-2.4 keV) of galaxy clusters rule out large quantities of cold
absorbing gas in these systems (e.g., Arabadjis \& Bregman 2000), our
spatial-spectral deprojection analyses of M87 in this paper and of
A1795 in PAPER1 strongly suggest that large quantities of warm
($\twarm\sim 10^6$ K) gas are distributed (at least) throughout the
central $\sim 100$ kpc of cluster cooling flows.

The total mass of the warm gas implied by the oxygen absorption is
consistent with the amount of matter deposited by an inhomogeneous
cooling flow (\S \ref{mass}). On the other hand, the mass deposition
profile (Figure \ref{fig.cf_tau}) and the profile of warm emission
fraction of the two-phase models (Figure \ref{fig.2t}) indicate that
the emission of the warm component is suppressed over $r\sim 1\arcmin
- 5\arcmin$ where the radio emission from the AGN jet clearly distorts
the X-ray isophotes \citep{owen}. This coincidence suggests that the
AGN has influenced the hot ISM in these central regions and may have
suppressed the cooling emission of the warm component (\S
\ref{feedback}; see Binney 1999). Within the central arcminute the gas
has apparently readjusted and is cooling while at large radii,
$r>5\arcmin$, the cooling flow was not disturbed significantly by the
AGN. A hybrid model of a standard cooling flow with AGN feedback seems
promising for M87.

It is puzzling, however, why the oxygen absorption optical depth does
not dip between $R\sim 1\arcmin - 5\arcmin$ as would be expected if
the absorption and excess soft emission arise from the same
material. Perhaps this is a result of the simplifications we have
employed in this investigation (e.g., see \S \ref{mass}), and with
rigorous consideration of the radiative transfer and the absorption
from several ionization states of oxygen (and carbon and nitrogen) a
self-consistent description of the multiphase medium will be obtained.

The highly significant detection of intrinsic oxygen absorption in the
\rosat PSPC data of M87 confirms the picture of a multiphase warm+hot
medium in cooling flows deduced from the lower S/N PSPC data of
galaxies and groups in PAPER1 and PAPER3. With new \chandra and \xmm
data the spatial distribution of the different phases can be mapped
with novel accuracy, and thus the structure and evolution of cooling
flows, and the role of AGN feedback, will be elucidated.

\acknowledgements

I thank D. Liedahl for discussions relating to the accuracy of plasma
emission codes and the anonymous referee for helpful comments.  This
research has made use of data obtained through the High Energy
Astrophysics Science Archive Research Center Online Service, provided
by the NASA/Goddard Space Flight Center. Support for this work was
provided by NASA through Chandra Fellowship grant PF8-10001 awarded by
the Chandra Science Center, which is operated by the Smithsonian
Astrophysical Observatory for NASA under contract NAS8-39073.

\end{document}